\documentclass[12pt]{emulateapj}
\usepackage{txfonts}
\usepackage{graphicx}

\newcommand{\kms}{${\rm km \; s^{-1}}$}
\newcommand{\ha}{H$\alpha$}
\newcommand{\hb}{H$\beta$}
\newcommand{\hg}{H$\gamma$}
\newcommand{\hd}{H$\delta$}

\shorttitle{Nebular Attenuation at $z=0.8$}
\shortauthors{Momcheva et al.}

\begin{document} 

\title{Nebular Attenuation in H$\alpha$-selected Star-forming Galaxies at $z=0.8$ from the \\ NewH$\alpha$ Survey}

\author{
Ivelina G. Momcheva\altaffilmark{1,2},
Janice C. Lee\altaffilmark{2,3,9,12},
Chun Ly\altaffilmark{2,3,4,10},
Samir Salim\altaffilmark{4},
Daniel A. Dale\altaffilmark{5},\\
Masami Ouchi\altaffilmark{2,6,7,9},
Rose Finn\altaffilmark{8},
and Yoshiaki Ono\altaffilmark{11}
}
\altaffiltext{1}{Astronomy Department, Yale University, New Haven, CT 06511, \email{ivelina.momcheva@yale.edu}}
\altaffiltext{2}{Carnegie Observatories, Pasadena, CA, 91101}
\altaffiltext{3}{Space Telescope Science Institute, Baltimore, MD 21218}
\altaffiltext{4}{Astronomy Department, Indiana University, Bloomington, IN 47405}
\altaffiltext{5}{Department of Physics and Astromomy, University of Wyoming, Laramie, WY 82071}
\altaffiltext{6}{Institute for Cosmic Ray Research, The University of Tokyo, Kashiwa, Chiba 277-8582, Japan}
\altaffiltext{7}{Institute for the Physics and Mathematics of the Universe (IPMU), TODIAS, The University of Tokyo, Kashiwa, Chiba 277-8583, Japan}
\altaffiltext{8}{Siena College, Physics Department, Loudonville, NY 12211}
\altaffiltext{9}{Carnegie Fellow}
\altaffiltext{10}{Giacconi Fellow}
\altaffiltext{11}{Department of Astronomy, Graduate School of Science, University of Tokyo, Japan}
\altaffiltext{12}{Visiting Astronomer, Spitzer Science Center, Caltech, Pasadena CA, 91125}

\begin{abstract}

We present measurements of the dust attenuation of
H$\alpha$-selected emission-line galaxies at $z=0.8$ from the
NewH$\alpha$ narrowband survey.  The analysis is based on deep follow-up
spectroscopy with Magellan/IMACS, which captures the strong rest-frame optical emission lines from
[OII]$\lambda$3727 to [OIII]$\lambda$5007.  The spectroscopic sample used in this analysis consists of 341 confirmed \ha\ emitters. We place constraints on the AGN fraction using diagnostics which can be applied at intermediate redshift. We find that  at least $5\%$ of the objects in our spectroscopic sample can be classified as AGN and 2\% are composite, i.e. powered by a combination of star-formation and AGN activity. We measure the dust attenuation for individual objects from the ratios of the higher order Balmer lines. The \hb\ and \hg\ pair of lines is detected with $S/N>5$ in 55 individual objects and the \hb\ and \hd\ pair is detected in 50 individual objects. We also create stacked spectra to probe the attenuation in objects without individual detections. The median attenuation at \ha\ based on the objects with individually detected lines is A(\ha)$=0.9\pm1.0$ magnitudes, in good agreement with the attenuation found in local samples of star-forming galaxies. We find that the $z=0.8$ galaxies occupy a similar locus of attenuation as a function of magnitude, mass and SFR as a comparison sample drawn from the SDSS DR4. Both the results from the individual $z=0.8$ galaxies and from the stacked spectra show consistency with the mass -- attenuation and SFR -- attenuation relations found in the local Universe, indicating that these relations are also applicable at intermediate redshift. 

\end{abstract}

\section{Introduction}

Accurate measurements of the dust attenuation in galaxies are critical for the determination of the intrinsic physical properties of galaxies, and are thus required for studies of galaxy formation and evolution.  In particular, dust attenuation affects key observables involving star formation and remains one of the largest sources of uncertainty in tracing the star formation rate (SFR) volume density over cosmic time \citep[e.g., ][]{reddy09}.

In the local Universe, a common way of measuring the nebular dust attenuation is to compare the observed ratios of the Balmer emission lines to their expected values in the absence of dust attenuation. 
The Balmer lines arise from the recombination and subsequent cascade transitions to the second ($n=2$) level of hydrogen. The expected relative fluxes of the Balmer lines can be computed theoretically \citep{menzelbaker37,bakermenzel38,seaton59,hummer98,osterbrock06,dopita}. These predicted line ratios are a function of the temperature and electron density in the nebular regions, however the functional dependence is weak and thus the line ratio variations are small ($\sim$ a few percent).  The attenuation caused by dust is wavelength-dependent, and it is stronger at shorter wavelengths. This wavelength dependence is described by the extinction law  \citep[e.g.,][]{dust, calzetti97}. The comparison between the predicted and observed line ratios (\ha/\hb\ for example) is then used to determine the reddening in the nebular regions via the extinction law. 

As a physically-motivated and easily-accessible attenuation diagnostic, Balmer decrements are widely employed in studies of the local Universe \citep[e.g.,][]{seaton79, gallagher89, calzetti97, jansen01, glazebrook03,moustakas06, moustakas06a,sullivan01,kewley01, kewley02a, kewley02b, moustakas10, finkelstein11, wij11a,wij11b}. The \ha\ attenuation, hereafter A(\ha) of normal, nearby,  star-forming galaxies can range from 0.1 to 1.8 magnitudes \citep[e.g.,][]{moustakas06a,lee09}, and is typically $\sim1$ magnitude in local galaxies similar to the Milky Way. Using star-forming galaxies from the Sloan Digital Sky Survey (SDSS), \citet{hopkins03} determine a median A(\ha) of 1.2 magnitudes based on the \ha/\hb\ decrement. Similarly, the SDSS study of \citet{brinchmann04} finds a mean A(\ha) of 1.0 magnitudes with a spread of 0.7 magnitudes for a SFR-weighted distribution and a mean of 1.3 magnitudes and a similar spread for a stellar mass weighted distribution.  Balmer decrements have also been used to calibrate a number of local empirical relations \citep[e.g.,][]{hopkins01,moustakas06,kennicutt09}, which can be used to estimate attenuation in the absence of more direct measurements . 
Such locally-calibrated relations have been applied to galaxy samples to $z\sim2$, under the assumption that the relations still hold at higher redshift \citep[e.g., ][]{dale10,ly11}. 

{ Clearly, however, the relation between attenuation and luminosity or mass may change over cosmic time. This expectation is based on the observation that the amount of dust in galaxies has been found to correlate with metal abundance \citep{heckman98, boissier04, asari07}.  The proposed explanation is that more metal-rich galaxies have higher dust to gas ratios and therefore higher extinction. The interrelatedness of SFR, stellar mass, metallicity and attenuation was also explored by \citet{garn10}, who show that the attenuation is most tightly correlated with stellar mass (more massive galaxies have built larger dust reservoirs), and the metallicity-attenuation relation is derivative through the mass-metallicity relation. Yet, the mass-metallicity relation has been shown to evolve with redshift \citep{savaglio05,erb06, maiolino08, mannucci09, mannucci10} such that galaxies of a given stellar mass show lower metallicities at earlier times. Therefore, the mass-attenuation and SFR-attenuation relations may also evolve in a similar way, and it is important to examine these relations with higher redshift datasets. }

For a direct comparison between low and high redshift we would ideally want to determine the attenuation consistently through the Balmer decrements. Until now, this has not been possible for a large sample. Studies of dust attenuation based on Balmer decrement measurements have largely been limited to galaxies in the local universe, primarily because at $z>0.4$ \ha\ shifts into the near infrared, making measurements of the H$\alpha$/H$\beta$ ratio challenging. Alternatively, the higher order Balmer lines \hb\ ($\lambda4861$), \hg\ ($\lambda4340$), \hd\ ($\lambda4101$), which are still observable in the optical up to $z=1$, may be examined.  However, observations of these lines are time-intensive since they are relatively weak. 


The few studies that examine Balmer decrements in intermediate redshift galaxies have mainly focused on samples which may not be representative of the overall star-forming population. For example, some studies focus on luminous and ultra-luminous infrared galaxies (LIRGs and ULIRGs, respectively) at $0.4<z<1.0$  \citep{hammer01,flores04,liang04}, and thus their findings of elevated nebular attenuation ($A_{V}=1.5$, 2.8 and 1.82, respectively) relative to typical local galaxies ($A_{V}=1.25$) may be expected. \citet{rodrigues08} focus on massive galaxies $\mathrm{M}_{\ast}>1.5\times10^{10}\mathrm{M}_{\odot}$ and find $A_{V}=1.53$ based on the \hg/\hb\ ratio, in good agreement with a complementary estimate based on a comparison between the infrared and H$\beta$ fluxes ($A_{V}=1.71$).  Such values are actually comparable to those found locally from the SDSS, when a similar range of galaxy masses is considered \citep{garn10}.  Finally, \citet{savaglio05} study 28 galaxies at $0.4<z<0.98$ from the K-band selected Gemini Deep Deep Survey \citep[GDDS, ][]{abraham04}.  The galaxies have stellar masses mostly between $10^{9}$ and 10$^{10}\ \mathrm{M}_{\odot}$, and may be a more representative sample of the star-forming population during this epoch.  However, when the H$\gamma$/H$\beta$ through H8/H$\beta$ were measured from a stacked spectrum of the galaxies, the line ratios yielded an attenuation of $A_{V}=2.13\pm0.32$, higher than what would be expected locally. 

In this paper, we use the higher order Balmer emission lines \hb, \hg, and \hd\ to constrain the amount of nebular dust attenuation in intermediate redshift star-forming galaxies.  Specifically, we wish to determine whether the average attenuation as a function of rest-frame optical luminosity, stellar mass and SFR changes significantly between the local universe and $z=0.8$.  This study improves upon previous work by using a large sample of \ha-selected galaxies within a narrow slice of redshift space down to SFR$\sim1 \mathrm{M}_{\odot}\ \mathrm{yr}^{-1}$ (i.e., comparable to typical SFRs in the local Universe) with deep optical spectroscopy which allows us to detect \hb, \hg\ and \hd. 
 The \ha\ sample selection and spectroscopic observations are described in \S 2. We estimate stellar masses for our $z=0.8$ sample using spectral energy distribution (SED) fits to the UV, optical and NIR photometry. These data-sets and the fitting procedure are described in \S 3.  In order to make a comparison between the attenuation in local and intermediate redshift galaxies, we select a sample of galaxies from the Sloan Digital Sky Survey (SDSS) in \S 4. Our emission-line selected sample is likely to contain both objects powered by star formation and active galactic nuclei (AGN). In \S 5 we identify AGN using a variety of techniques and place a lower limit on the AGN fraction in our sample. The main dust attenuation analysis is presented in \S6. We present Balmer decrement measurements for individual objects at z=0 and z=0.8 as a function of J-band magnitude, mass and SFR in \S 6.1, \S 6.2 and \S6.3. However, individual detections are only available for a biased portion of the $z=0.8$ sample. To remedy that, we create stacked spectra from the full spectroscopic sample in bins of magnitude, mass and \ha$+$[NII]$\lambda$6584 flux. The Balmer decrements measured from the stacked spectra are presented in \S 6.4. In \S 7 we compare our results with other local and intermediate redshift studies, discuss them in the context of galaxy evolution and examine the limitations and biases in our sample. Overall we find that the $z=0.8$ galaxies occupy the same locus of attenuation as a function of magnitude, mass and SFR. The mean nebular attenuation we find is A(\ha)$\sim1$ magnitude. The stacked spectra also indicate a mean attenuation of $\sim1$ magnitude. The results from the stacked spectra are consistent with local mass-attenuation and SFR-attenuation relations, suggesting that these relations can also be used in the intermediate redshift Universe. We summarize our findings and discuss future work in \S 8. 

\section{IMACS Spectroscopy}

The galaxy sample used in this analysis consists of 341 \ha\ emitters in a narrow slice of redshift centered at $z=0.8$.  Deep follow-up spectroscopy of these galaxies was obtained with the IMACS instrument on the Magellan-I 6.5-m telescope. Here we give an overview of the initial near-IR narrow-band imaging used to select the sample of H$\alpha$ emitters, and then describe the IMACS follow-up spectroscopic observations and their reduction.

\subsection{The NewH$\alpha$ Survey: Narrowband Imaging and
Emission-Line Galaxy Candidate Sample}

The New\ha\ Survey is a program which has obtained \ha\ selected samples at intermediate redshift \citep{lee2012a,lee2011}. Our program has been structured to efficiently capture statistical samples of both luminous (but rare) and faint emission-line galaxies.  We do this by combining the near-infrared imaging capabilities of NEWFIRM at the KPNO/CTIO 4 m telescopes (FoV 27\farcm6 x 27\farcm6) to cover large areas, and FourStar at the Magellan 6.5m (FoV 10\farcm9 x 10\farcm9 arcmin) to probe luminosities that are about a factor of three deeper over smaller areas. For both cameras, we have designed a pair of 1\% narrowband filters which fit within high atmospheric transmission, low OH airglow windows at 1.18 $\mu$m and 2.09 $\mu$m.  With these two filters we obtain deep H$\alpha$-selected galaxy samples at $z$=0.8 (near the beginning of the ten-fold decline in the cosmic SFR density) and at $z$=2.2 (near the peak of the cosmic SFR density).

The work presented in this paper focuses on H$\alpha$ emitters at $z$=0.8, which are detected in NEWFIRM narrowband 1.18 $\mu$m (hereafter NB118) and $J$ imaging of a 0.82 deg$^{2}$ region in the Subaru-XMM Deep Survey \citep[SXDS,][]{furusawa08}, and a 0.24 deg$^{2}$ region in the SA22 field \citep[a.k.a., SSA22,][]{sa22}.
Here we summarize the NB118 observations, data reduction, and selection method used to produce samples of emission-line galaxy candidates that are targeted for IMACS spectroscopy.  Further details on the NEWFIRM NB118 imaging component of the survey can be found in \citet{ly11} and \citet{lee2011}.

Three NEWFIRM pointings in the SXDS ($\alpha=02^{h}18^{m}$; $\delta = -05^{\circ}00\arcmin$) and one in SA22 ($\alpha=22^{h}17^{m}$; $\delta = -00^{\circ}16\arcmin$) were obtained in 2007 December, 2008 September and 2008 October at the KPNO 4m telescope.  The cumulative exposure times for each pointing ranged from 8.47 to 12.67 hr in NB118 and from 2.40 to 3.97 hr in $J$. The median seeing during our observations was $\sim1\farcs2$, and varied between 1\farcs0 and 1\farcs9, so point sources are adequately sampled by NEWFIRM's 0\farcs4 pixels. Standard near-infrared deep field observing procedures and reduction techniques were followed. The 3$\sigma$ limiting magnitudes, in apertures containing at least $\sim$80\% of the flux of a point source, range from 23.7$-$24.2 AB magnitudes in NB118 and 23.4$-$24.1 in $J$.  

Sources are selected as emission-line galaxy candidates if they show a $J-$NB118 color excess which is significant at the 3$\sigma$ level and is greater than 0.2 mag.  The minimum of 0.2 mag is based on the scatter in the color excess for bright point sources. Coarse corrections for the continuum slope are applied based on the $z^\prime-J$ color,
using publicly available Subaru/Suprime-Cam $z^\prime$ data \citep{furusawa08} where available.  The overall procedure follows general selection techniques commonly used in narrow-band surveys \citep[e.g.,][]{fujita03,ly07,shioya08,villar08,sobral09}. A total sample of 937 emission-line galaxy candidates meeting these criteria 
was obtained over the four NEWFIRM pointings. A catalog, which contains emission line fluxes and equivalent width measurements of these sources, is provided in \citet{lee2011}.

\subsection{Spectroscopic Observations}

Deep follow-up spectroscopy of the New\ha\ NB118 excess sample was performed in 2008-2009 with the Inamori Magellan Areal Camera and Spectrograph \citep[IMACS;][]{dressler06} at the 6.5m Magellan-I telescope. IMACS enables multi-object spectroscopy with slit masks over a 27\farcm4 diameter area (an excellent match to the field-of-view of NEWFIRM), and has good sensitivity to $\sim$9500\AA. These two characteristics make IMACS an ideal instrument for optical spectroscopic follow-up of New\ha\ NB118 excess sources, and in particular, \ha\ emitters at $z=0.80$. Our observational set-up yields spectral coverage from 6300\AA\ to 9600\AA\ (corresponding to rest-frame $\sim$3500 to $\sim$5300\AA), and captures the strong rest-frame optical emission lines from [OII]$\lambda$3727 (observed at 6720\AA) to [OIII]$\lambda$5007 (observed at 9030\AA) for galaxies at $z=0.8$. 

The spectroscopy used in this analysis was obtained over three
observing runs in 2008 October, 
2009 September and 2009 November, and was mainly focused on sources in the SXDS field.   Observations in the SA-22 field were carried out as time allowed, when the SXDS field was not visible. The spectra were taken with the f/2 camera and the Red 300 l/mm grating\footnote{There are two 300 l/mm gratings. The blue one has a blazing angle of 17.5 degrees and a central wavelength of 6700 \AA. The red one has blazing angle of 26.7 degrees and a central wavelength of 8000 \AA. We used the latter.}, which has a dispersion of 1.25 \AA/pix.  Masks were designed using 7\arcsec$\times$1.5\arcsec\ slits.  This set-up results in an observed resolution of 9\AA, and corresponds to 5 \AA\ in the rest-frame for the $z=0.8$ objects. The WB6300-9500 blocking filter was used to limit the length of each spectrum on the detector and allow more targets to be observed simultaneously. An atmospheric dispersion corrector was used in all observations. 

Sources were also observed with the same set-up during 2008 November and 2008 December as mask fillers for a program targeting high redshift Ly$\alpha$ emitters (PI Masami Ouchi). The majority of these observations used a set-up identical to ours, but for one mask, a different blocking filer, WB4800-7800 was used instead of WB6300-9500.  The use of the WB4800-7800 filter results in a rest-frame spectral coverage of 2670-4200 \AA\ range for galaxies at z=0.8.  The impact on our overall spectroscopic follow-up program is minimal however, as this particular mask only included 35 NB118 excess emitters, eight of which show a redshift of z=0.8.
Of these eight, five were observed again on other masks, leaving only three objects without spectroscopic coverage to the red of 4200 \AA. 

Nightly observing procedures were as follows.
Bias frames were taken before the start of the night. 
The slit mask was aligned on the field using at least 10 bright alignment stars.  
The exposure times for individual science frames  were 20 or 30 minutes. 
The alignment was checked every hour and the mask was re-positioned as needed.
Comparison arc lamp spectra and flat field exposures were also taken every hour throughout the night. The arcs were observed with one set of He, Ne, Ar lamps, and
flat frames were taken with the quartz lamps. 
At least one spectrophotometric standard was observed at the beginning of the night, half-way through and/or at the end of the night, as time allowed, when conditions were photometric. 
The standard stars were selected to be close in air-mass to the science target fields from the ESO catalog\footnote{\url{http://www.eso.org/sci/observing/tools/standards/spectra/stanlis.html}}. The standard stars were observed with a 1.5\arcsec\ long slit mask. Arc and flat observations with the long slit mask immediately followed. 

Overall, IMACS spectroscopy was obtained for 515 
out of 937 objects in the NEWFIRM 3$\sigma$-selected NB118 excess catalog (386/661 in SXDS and 127/276 in SA22).  In addition, 214 SXDS objects which showed color excesses of less than 3$\sigma$ significance were also observed to help test the completeness of the sample.  Spectra were taken through 15 different slit masks covering 0.62 sq. deg in the SXDS (4 of the masks were from the Ouchi et al. high redshift Ly$\alpha$ program), and through 3 masks covering 0.24 sq. deg in SA22.  Targets in the SXDS were prioritized for inclusion on the masks (via the IMACS mask design software) based on their photometric redshifts \citep{furusawa11}. Highest priority was given to sources likely to be intermediate redshift emission-line galaxies ($z_{phot}>0.7$), while low redshift sources ($z_{phot}<0.7$) were considered to be low-priority. Slits were placed on 80 to 96\% of high-priority objects. The remaining high-priority objects were not observed due to slit conflicts and generally lie in areas with high density of targets. Photometric redshifts for the SA22 field were not available to us at the time that the masks were designed, and no prioritization was done for the NB118 excess emitters in that field. 

\begin{figure}
\figurenum{1}
\epsscale{1.0}
\plotone{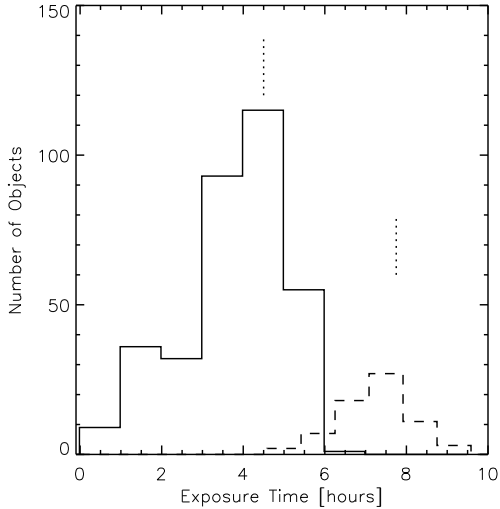}
\caption{Distribution of the cumulative spectroscopic exposure times for objects in the sample  of 341 \ha-selected galaxies at $z=0.8$ which are used in this analysis (solid line histogram). Also shown are the cumulative integration times for objects with which were observed during more than one observing run and then co-added (dashed line histogram). The medians are 4.5 and 7.75 hours respectively (dotted vertical lines).\label{exptime}
}
\end{figure}

The cumulative integration times per mask range between one and seven hours, with a median of 4.5 hours. Figure \ref{exptime} shows the distribution of cumulative integration times for the spectroscopically confirmed \ha\ emitters that are the focus of this analysis. 
For a sample of 140 NB118 excess sources we obtained observations during multiple observing runs. The majority of these objects were observed in two runs, however 15 of them were observed during three or more runs. Most of these observations were incidental duplications except for a sample of 23 $z=0.8$ H$\alpha$ emitters in the SXDS-South field, which were intentionally re-targeted to build up signal in objects with detected Balmer lines which had initial S/N$<5$. 
The cumulative exposure times for the H$\alpha$ emitters which were observed during multiple runs range from 5.5 to 11.5 hours, with a median of 7.75 hours (Figure \ref{exptime}). 

\begin{deluxetable}{lrrr}
\tablecolumns{4}
\tablewidth{0pc}
\tabletypesize{\footnotesize}
\tablecaption{Sample Summary \label{tab:sample}}
\tablehead{
\colhead{ } & \colhead{SXDS} & \colhead{SA22} & \colhead{Total}
}
\startdata
3$\sigma$ NB118 emitter & 661 & 276 & 937 \\
3$\sigma$ NB118 emitters targeted with IMACS & 386 & 127 & 515 \\
Other objects targeted with IMACS\tablenotemark{a}  & 214 & \nodata & 214\\
Confirmed \ha\ emitters with IMACS\tablenotemark{b} &299 & 42 & 341 \\
\ha\ emitters with mass estimate & 274 & \nodata & 274\\
\hline
Objects in spectral stacks & & &174 \\
H$\beta$, H$\gamma$ detected at $\geq5\sigma$ & & &55 \\
 H$\beta$, H$\delta$ detected at $\geq5\sigma$ & & & 50 \\
 H$\beta$, H$\gamma$, H$\delta$ detected at $\geq5\sigma$ & & & 34 \\
\enddata
\tablenotetext{a}{These are candidate emitters with $<3\sigma$ significance.}
\tablenotetext{b}{Objects at $0.78<z<0.83$, independent of their NB118 detection significance.}
\end{deluxetable}

\subsection{Spectroscopic Reduction}\label{specred}

The IMACS data were reduced using 
the instrument-specific, publicly-available COSMOS software, developed by the IMACS instrument team at Carnegie Observatories \citep{oemler}. COSMOS relies on an accurate optical model of the spectrograph to determine the expected positions of the spectral features in the detector plane. The data reduction includes the following steps: (1) an approximate wavelength solution for each slit is established using three bright, isolated arc lines; (2) the wavelength mapping is adjusted using all bright arc lines; the mapping is done separately for each arc exposure throughout the night; (3) bias frames are created by averaging ten bias exposures taken at the beginning of each night; (4) flat-fielding frames are created using each set of quartz frames throughout the night; (5) bias subtraction and flat-fielding are carried out for each science frame; (6) sky-subtraction is carried out on all science frames; (7) each 2-dimensional science spectrum is extracted to an individual image and its wavelength linearized; (8) the separate exposures are co-added. Each science frame is paired with the wavelength map and flat field closest in time, typically preceding or following the science exposure. The one dimensional spectra are then extracted using an optimal extraction routine \citep{optimal} which aims to maximize the signal-to-noise of the resulting spectrum. An error spectrum is also produced. Heliocentric corrections were calculated for each mask at the midpoint of the observing night and added to the headers.

\begin{figure*}
\label{sky}
\figurenum{2}
\epsscale{1.0}
\plotone{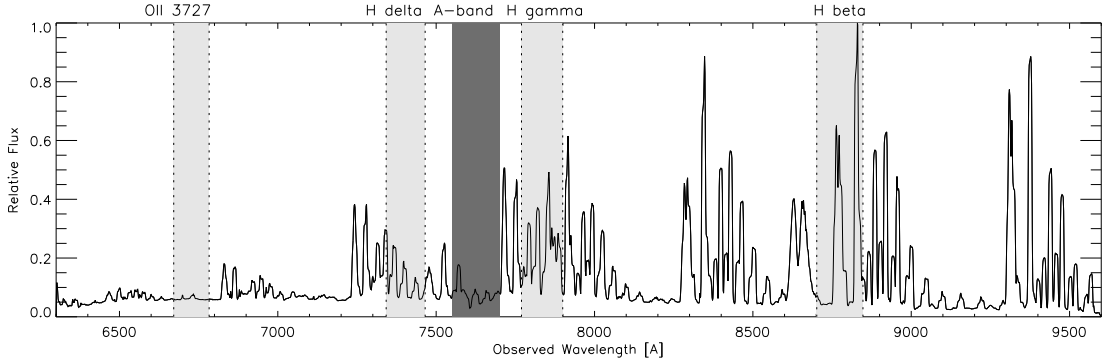}
\caption{Sky spectrum in the wavelength range of our science spectra. The positions of the Balmer lines for redshifts between 0.79 and 0.83, the redshift window for which the parent NEWFIRM near-IR narrow-band observations are sensitive to \ha, are shaded light gray. The position of the atmospheric absorption A-band is shaded dark gray.}
\end{figure*}

Figure 2 illustrates the sky spectrum over the wavelength range of our observations, indicating the expected positions of the Balmer emission lines over the redshifts range of our narrowband selected \ha\ sample. The figure makes it clear that obtaining good sky subtraction is critical for accurate line measurement. The COSMOS software employs a two-dimensional sky subtraction algorithm \citep{kelson}. We vary 
parameters involving the fitting of the sky background
to minimize the sky residuals. The sky subtraction is done by fitting a second order spline to the sky with a 0.75 pixels knot spacing.  We also exclude the central 8 pixels of the slit from the fit to avoid subtraction of the emission lines in the science objects. The use of the temporally closest flat-field frame also reduced the residuals because it minimized the differences between the low-level fringing patterns on the flats and the science frames. Sky residuals were challenging to minimize to the red of 8500 \AA.
For this reason,
we have visually inspected the spectra of all objects included in our work and excluded $\sim100$ spectra (out of the 341) which are dominated by sky subtraction errors.

The standard star spectra were reduced in the same manner as the science frames. We created a sensitivity function for each night using the noao.onedspec.calibrate and noao.onedspec.sensfunc routines in IRAF\footnote{IRAF is distributed by the National Optical Astronomy Observatories, which are operated by the Association of Universities or Research in Astronomy, Inc., under cooperative agreement with the National Science Foundation.}. If multiple standard stars were observed on a given night, they were all combined to produce a single, average sensitivity function. The sensitivity function was then applied to all science spectra to produce flux-calibrated spectra. This flux calibration, however, is not absolute because it does not account for slit losses. We side-step this issue by using line ratios in the analysis. However, we have estimated the effect of slit losses and offsets in the flux calibration by synthesizing magnitudes from the spectra and comparing them to available photometry from Subaru/Suprime-Cam in the medium bandpasses NB816 and NB921. We find that the spectra are missing $\sim10\%\pm10\%$ of the continuum flux relative to the magnitude measured in a 2\arcsec\ aperture (for $\sim$1\arcsec\ PSF).

Redshifts were determined using an automated cross-correlation routine based on the SDSS redshift-finding algorithm \citep{cool08}, customized to accept the IMACS spectra. The routine uses a $\chi^{2}$ minimization to compare each observed object to a library of galaxy and QSO templates that are linear combinations of eigen-spectra. Each redshift was 
verified by visually examining the 20 best $\chi^{2}$ fits as well as the sky-subtracted, two-dimensional spectrum to confirm all emission features. For 102 objects which were observed 
during different runs, each epoch of observation yielded a redshift measurement. We use these repeat redshift measurements to 
estimate the redshift error of the spectroscopic catalog. The mean difference between the two redshift measurements is $-6.3\times10^{-6}$ ($-2$ km s$^{-1}$) with a standard deviation of $4.9\times10^{-4}$ ($146$ km s$^{-1}$).  

Of the 729 objects from the 3$\sigma$-selected NB118 excess catalog targeted for IMACS spectroscopy, 225 (42) in the SXDS (SA22) have redshifts between 0.78 and 0.83, which confirms that the narrowband photometric excess is due to H$\alpha$ emission.  For 40\% of the NB118 excess sample in the SXDS without spectroscopy, H$\alpha$ emitters are identified using a color classification scheme based on publicly available Suprime-Cam $R_c i^{\prime} z^{\prime}$ photometry, which is calibrated using the galaxies with IMACS redshifts \citep[see Figure 4 in ][]{ly11}.  An additional 110 SXDS sources are identified as H$\alpha$ emitters in this manner.  Thus, IMACS spectra are available for 67\% of the SXDS $z=0.8$ H$\alpha$ sample.  In the analysis that follows, 74 SXDS sources which have $0.78<z<0.83$, but which show an NB118 excess at a significance lower than the 3$\sigma$ cut-off are also included.  A final color selection to identify the H$\alpha$ emitters among the SA22 NB118 excess sources without spectroscopy has not yet been performed, but the 42 confirmed SA22 H$\alpha$ emitters are included in the analysis as well to increase the size of the sample. 

\subsection{Spectral Fitting}

\begin{figure*}
\figurenum{3a}
\label{fig:3a}
\epsscale{0.8}
\plotone{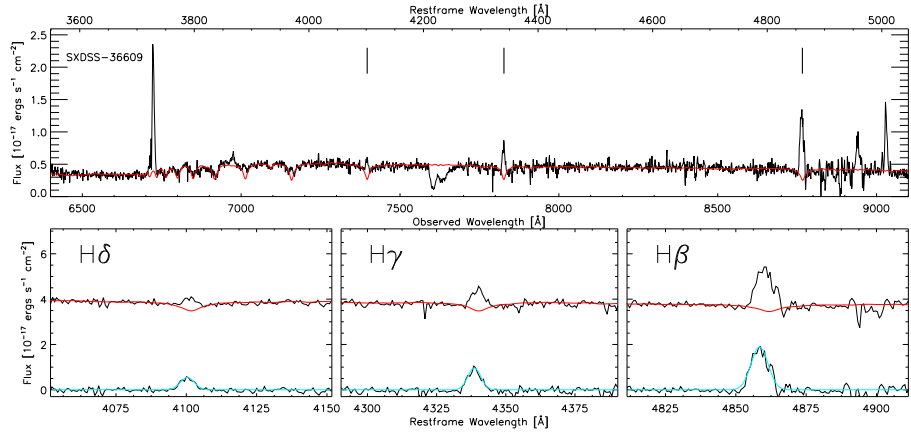}
\caption{A representative spectrum of a z=0.8 \ha-emitter in the SXDSS-36609, which has one of the highest \hb\ fluxes in the sample. This object is also identified as composite star-forming and AGN galaxy by the MEx and Spectral type methods (see Section 5). {\it Top:} The observed spectrum is shown (black) together with the continuum fit (red) as a function of the observed (bottom axis) and the restframe (top axis) wavelengths. Tick marks above the spectrum indicate the positions of the \hd, \hg, and \hb\ lines. {\it Bottom: } A close-up of the Balmer lines \hd, \hg, and \hb\  in the observed spectrum with the continuum fit (black and red lines) and in the continuum-subtracted spectrum fitted with a Gaussian (black and blue lines).}
\end{figure*}

\begin{figure*}
\figurenum{3b}
\label{fig:3b}
\epsscale{0.8}
\plotone{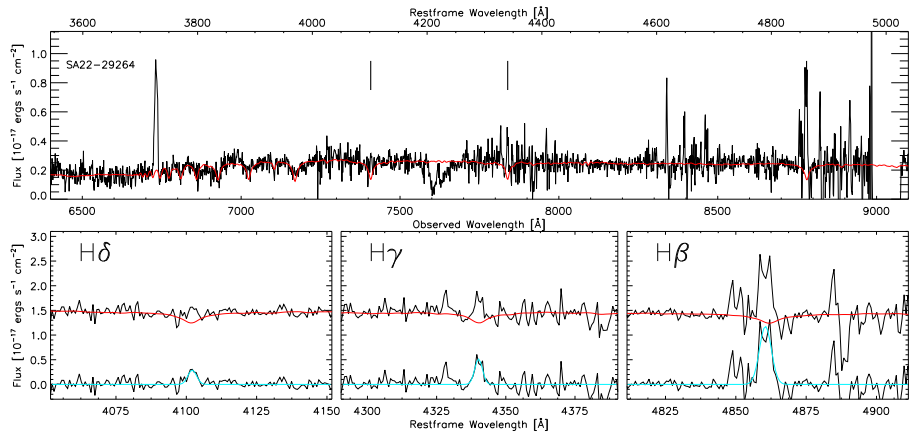}
\caption{Same as Figure \ref{fig:3a} but for the \ha\ emitter SA22-29264, which has a mean \hb\ flux. }
\end{figure*}

\begin{figure*}
\figurenum{3c}
\label{fig:3c}
\epsscale{0.8}
\plotone{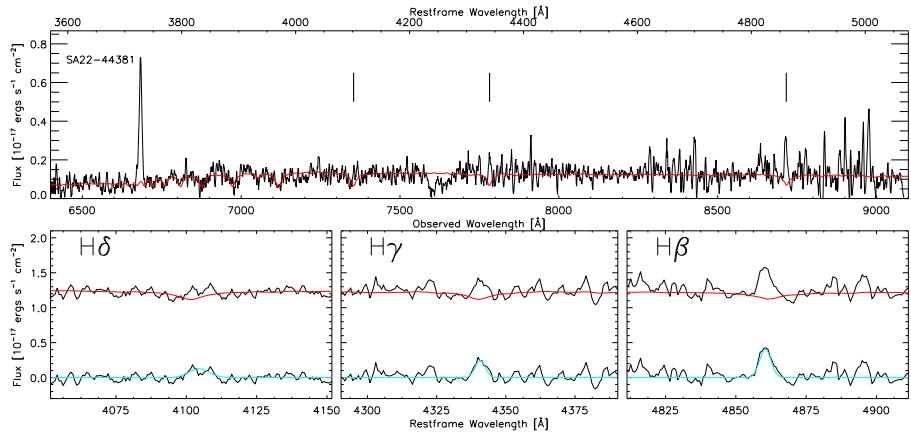}
\caption{ Same as Figure \ref{fig:3a} but for the \ha\ emitter SA22-44387, which has one of the lowest \hb\ fluxes. The \hd\ emission line is not detected. }
\end{figure*}

In order to obtain continuum-subtracted line flux measurements, self-consistently corrected for stellar absorption, the fluxed spectra were fit with stellar population models using a routine similar to that developed by \citet{christy} for SDSS. This fitting method assumes that any galaxy 
star formation history (SFH) can be approximated by a sum of discrete bursts and thus the spectra are fit by linear combinations of single burst templates.  The template library is generated from a linear combination of \citet{bc03}  models assuming solar metallicity and Chabrier initial mass function \citep[IMF, ][]{chabrier03}. We use templates with ten different ages - 10 Gyr, 5 Gyr, 2.5 Gyr, 1.43 Gyr, 0.90 Gyr, 0.64 Gyr, 0.29 Gyr, 0.10 Gyr, 25.1 Myr, and 5.01 Myr. We shift each galaxy spectrum to the rest frame and then construct the best fitting model, convolved to the proper spectral resolution. Internal velocity dispersion of 80 \kms\ is assumed. The fit includes correction for the foreground extinction using the \citet{schleg} extinction map and the \citet{dust} Milky Way extinction curve. The intrinsic reddening is treated as a free parameter using the \citet{cf_2000} law. The continuum fit is subtracted from the spectrum, and the emission line fluxes and equivalent widths are measured. The velocity offset of the emission lines from the systemic redshift is assumed to be the same for all lines. The line widths are also assumed to be the same.  

In the integrated spectra of all but the most vigorously star forming galaxies, 
the Balmer emission lines typically appear to be superimposed on top of the absorption lines of the stellar continuum and partially fill them (e.g., Figure 3).  Hence, an absorption correction must be applied to recover the correct Balmer emission-line flux.
The continuum fitting and subtraction procedure provides a self-consistent absorption correction to all emission lines. We check whether these corrections are reasonable, given other values quoted in the literature.
We measure the equivalent widths of \hb, \hg\ and \hd\ in the continuum fits and find mean values of 3.1, 2.2 and 3.4\AA, respectively. Similar values have been measured for local and intermediate redshift samples: \citet{kobulnicky99} find that for their sample of local galaxies the absorption correction of \hb\ is between 1 and 6\AA, with a mean of 3$\pm2$\AA; \citet{miller02} adopt a morphology dependent correction assuming that the corrections at \hb, \hg\ and \hd\ are identical and range from 1.2 to 4.1 (Sa to Irr); and \citet{savaglio05} determine corrections of 3.6\AA\ and 3.4\AA\ for \hb\ and \hg, respectively, based on a composite spectrum of $0.4<z<1.0$ galaxies. 

\begin{figure*}
\figurenum{4}
\label{fig:linedistr}
\plotone{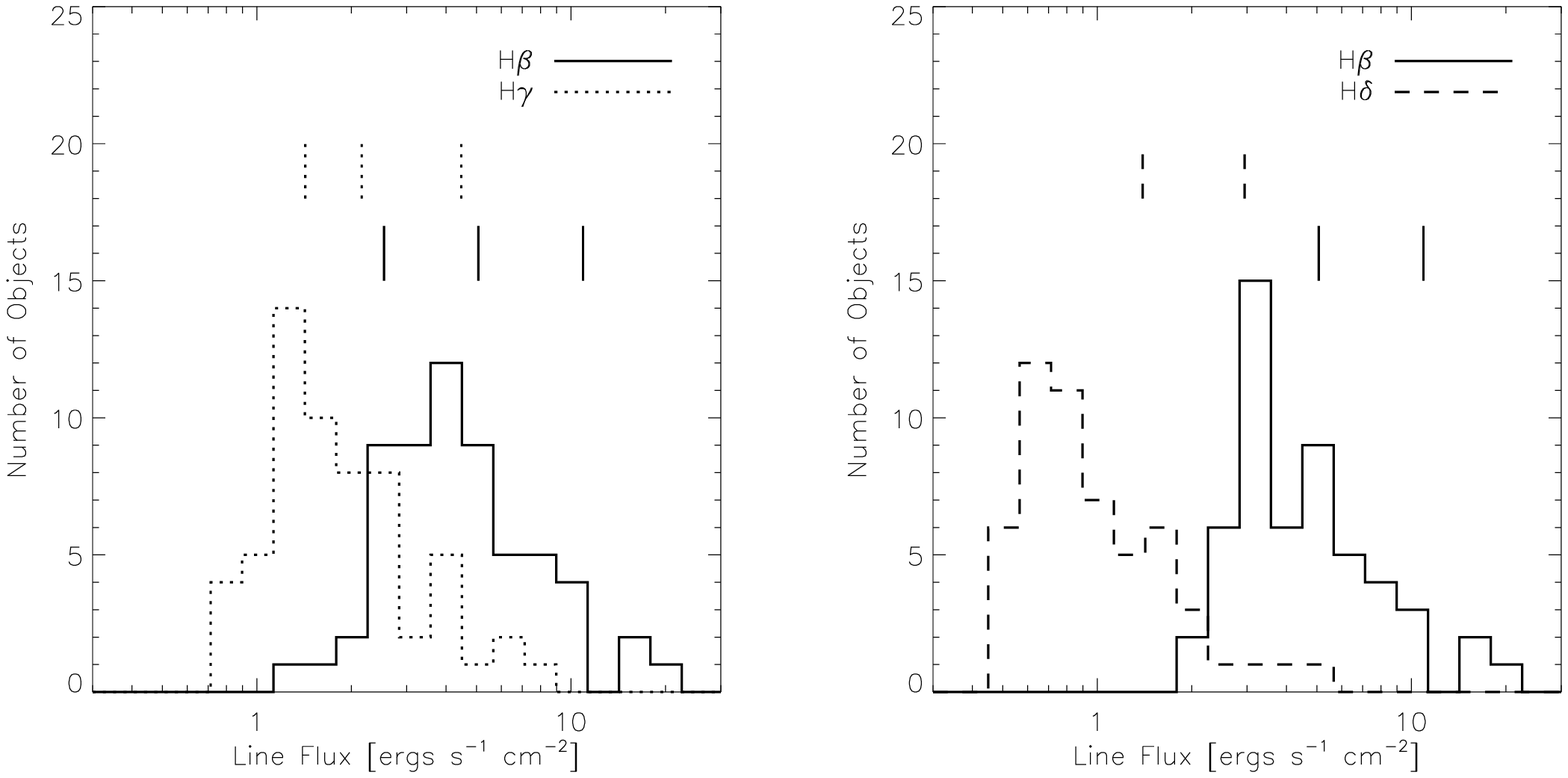}
\caption{Distributions of the fluxes of the Balmer lines in the samples with individual line detection after the absorption correction. {\it Left: } The \hb\ (solid line) and \hg\ (dotted line) emission lines are simultaneously detected with S/N$>5$ in 55 objects. The vertical lines above the distributions indicate the \hb\ and \hg\ fluxes of the objects shown in Figures 3a, 3b, and 3c. {\it Right: } The \hb\ and \hd\ (dashed lines) emission lines have been simultaneously detected with S/N$>5$ in 48 objects. The vertical lines above the distribution indicate the \hb\ and \hd\ fluxes of the objects shown in Figures 3a and 3b. \hd\ is not detected in the objects shown in Figure 3c. The line fluxes presented here have not been corrected for foreground attenuation.}
\end{figure*}

In Figures 3a, 3b  and 3c we show three representative spectra from our sample. The spectra are ordered by decreasing \hb\ flux. Figure 3a presents an object with a high \hb\ flux, also identified as a composite SF+AGN object (see Section 5 ), Figure 3b is an object with a mean \hb\ flux, and Figure 3c is an object with a low \hb\ flux. The three objects all have the \hb\ and \hg\ lines detected with a S/N$>5$ and are included in the \hg/\hb\ decrement sample considered later in this paper. While the sky line residuals are visible in the spectra, the Balmer lines are clearly detected. The continuum is also well detected which allows for a successful continuum fit and reliable subtraction of the stellar absorption in the Balmer lines. The distributions of \hb, \hg\ and \hd\ line fluxes are presented in Figure \ref{fig:linedistr}. The \hb\ and \hg\ lines have been simultaneously detected in 55 objects with S/N$>5$ (left panel). The \hb\ and \hd\ lines have been simultaneously detected in 49 objects with S/N$>5$ (right panel). The line fluxes of the objects in Figures 3a, 3b and 3c are indicated above the distributions. We note that these fluxes are absorption corrected (as described above), but no corrections for slit losses or attenuation have been applied. 

\section{Stellar Masses}

In the local Universe it has been shown that there is a correlation between stellar mass and dust attenuation for star-forming galaxies \citep[e.g.,][]{brinchmann04,garn10}. Such a correlation may result if star-formation and dust-formation are linked such that the build-up of stellar mass leads to an increase in dust content and attenuation. Using our dataset we can investigate whether any such relationship holds at $z=0.8$. To carry out this analysis, stellar masses for the spectroscopically observed galaxies must be computed. Stellar masses can be reliably obtained with spectral energy distribution (SED) fits to multi-wavelength photometry \citep[e.g.,][]{salim07,walcher}. 
A wealth of deep multi-wavelength imaging is available for the SXDS, 
and we use these data to derive stellar masses for the spectroscopically confirmed H$\alpha$ emitters in that field.  Specifically, we use the publicly available Subaru Suprime-Cam SXDS optical dataset \citep{furusawa08}, combined with the NEWFIRM J-band photometry and new GALEX NUV observations obtained by the NewH$\alpha$ team (P.I. S. Salim, GI6-005).  Comparable data are not available for the SA22 field, so we confine the portions of the analysis involving stellar mass to the SXDS H$\alpha$ sample. 
In this section we describe the NUV and optical observations, and the SED fitting procedure.  The J-band imaging is described in \citet{ly11} and \citet{lee2011}, and is also summarized in Section 2.1 above.

\subsection{Optical Photometry}

Deep optical imaging in the SXDS over 1.22 deg$^{2}$ in five contiguous fields has been obtained with Suprime-Cam on the Subaru 8.2-m telescope as part of the Subaru Telescope Observatory Projects \citep{furusawa08}.
Imaging has been performed in five broadband filters, $B$, $V$, $R_{c}$, $i'$, and $z'$, to the depths of $B=28.4$, $V=27.8$, $R_{c}=27.7$, $i'=27.7$, and $z'=26.6$, respectively (AB, $3\sigma$, 2\arcsec aperture).
The three fields in the SXDS observed with NEWFIRM are roughly coincident with the Suprime-Cam SXDS-S, SXDS-W and SXDS-N pointings \citep[Fig. 2,][]{nakajima11}.  The reduction of these data is described in \citet{furusawa08}. 

\subsection{GALEX Data}

Near-UV (NUV, $\lambda=2316$ \AA) photometry was obtained from deep imaging with the Galaxy Evolution Explorer \citep[GALEX, ][]{martin05,morrissey07}. At $z=0.8$, the NUV band samples the rest-frame far-UV, providing strong constraints on the star formation rate and history, and additional leverage on the stellar mass-to-light ratio and the stellar mass. Our NEWFIRM fields in the SXDS partially overlap with three archival GALEX ``tiles" ($1.2\deg$ diameter circular fields) which have exposure times 28, 26 and 30 ks.  To increase the overlap and the effective depth, we acquired one additional GALEX tile with an exposure time of 46 ks.  
The mean 5 sigma depth for the combined NUV imaging in the NEWFIRM fields is 25.3 AB (exposure time $\sim$80 ks). This is only $\sim$1 magnitude shallower than the deepest NUV tile \citep[the EGS field; ][]{salim09}, which, as of GALEX Release 6 (``GR6") has a total integration time of 257 ks.

At $z=0.8$, sources are unresolved in the GALEX imaging (the GALEX FWHM is $5\arcsec$) and are subject to blending. The GALEX pipeline catalogs, based on aperture photometry, do not optimally deal with these issues. Therefore we carried out a custom PSF source extraction using DAOPHOT. We set the DAOPHOT detection and photometry parameters to optimize the completeness of the detections, and
 currently generate separate flux catalogs for each of the four GALEX tiles. Each source list is matched to the NEWFIRM source list using a $3\arcsec$ search radius, which corresponds to $\sim3\sigma$ positional uncertainty. Of the 299 spectroscopically confirmed \ha\ sources at $z=0.8$, 85\% are detected in at least one of the NUV tiles. 
The photometry from the four source lists is then averaged
for objects detected in more than one tile, where the
average is weighted by photometry errors.
More details on the GALEX data available in the SXDS field is provided by \citet{salim_prep} and \citet{lee2011}. \citet{salim_prep} will also present a catalog of the de-blended photometry, as measured from the stacked SXDS mosaic, and further optimization of the NUV photometry is in progress.

\subsection{SED Fitting}

Stellar masses are estimated by fitting SEDs constructed from the multi-wavelength photometry described above\footnote{This SED fitting is distinct from the fitting of the IMACS spectra discussed in Section  \ref{specred}, which provided restframe optical ($3700--5200$\AA) continuum levels for the continuum subtraction. The SED fitting described in this section is based on photometry from the NUV to the NIR.}.  The SEDs are fit with a grid of BC03 stellar population synthesis models. The methodology is identical to that of \citet{salim07, salim09} and we refer the reader to these papers for a detailed discussion. Model libraries are built with a wide range of star formation histories and metallicities. Only models with formation ages shorter than the age of the universe at a given redshift are allowed. Each model is attenuated according to the prescription of \citet{cf_2000}, with variable values of both the total optical depth and the fraction of the total optical depth due to attenuation of the ambient ISM ($\tau_{V}$ and $\mu$, respectively). The dust attenuation is heavily constrained by the UV slope which gets shallower with increasing attenuation \citep{calzetti94}. However, differences in the star formation histories can produce significant scatter between the UV slope and dust attenuation \citep{kong04}. This is overcome in our modeling because the NIR and optical data help to constrain the age.  Intergalactic reddening is included via the prescription of \citet{madau96} and a Chabrier IMF is assumed.
The spectroscopic redshift provides the luminosity distance which
allows the apparent model quantities to be scaled to absolute values. 

The SED fitting performed in this study involves up to six flux points (NUV, $B,V, i, z, J$), their photometric errors and the spectroscopic redshifts.  For the spectroscopically confirmed $z=0.8$ H$\alpha$ emitters, only 20\% of the sample is fit with fewer than six flux points, and a minimum of four points are used.  The $r$\arcmin-band photometric points are excluded for the $z=0.8$ sample in order to avoid the restframe 3700\AA\ discrepancy with the stellar synthesis models \citep{salim09}. For each galaxy the observed flux points are compared to the model flux points and the goodness of fit ($\chi^{2}$) determines the probability weight of a given model. The average of the probability distribution of each parameter (e.g., stellar mass) is the nominal estimate of that parameter and its width is used to estimate the errors and confidence intervals.

The final sample of $z=0.8$ H$\alpha$ emitters with determined masses is 274 objects. 
The majority of galaxies in the sample are well fit and the median $\chi^{2}$ per degree of freedom of the best fitting SED models is one.  
Seven objects are excluded from the sample due to poor fits (i.e., if $\chi^{2}$ of the best fitting model is $>$10).  
Of these seven, one is a broadline AGN and one is a possible gravitational lens with two sets of emission lines at different redshifts.  The remaining five poor SED fits are likely due to errors in the photometry. 

Stellar mass uncertainties for the $z=0.8$ H$\alpha$ sample 
are on average 0.10 dex (25\%, $1\sigma$ confidence interval) and are determined from the widths 
of the probability distribution functions for each galaxy.
These estimates include uncertainties from input photometry 
and parameter degeneracy (e.g. with respect to SFH and dust). 
Additional systematic uncertainties may arise from the models 
themselves and the choice of IMF \citep[e.g., ][]{maraston05, conroy09, taylor11}. \citet{salim05} show that the masses derived via the method used here are in good agreement with those of \citet{kauffmann03} and the scatter in the differences between the two mass estimates is well matched by the uncertainties in the two studies.

\begin{figure}
\figurenum{5}
\label{}
\epsscale{1.0}
\plotone{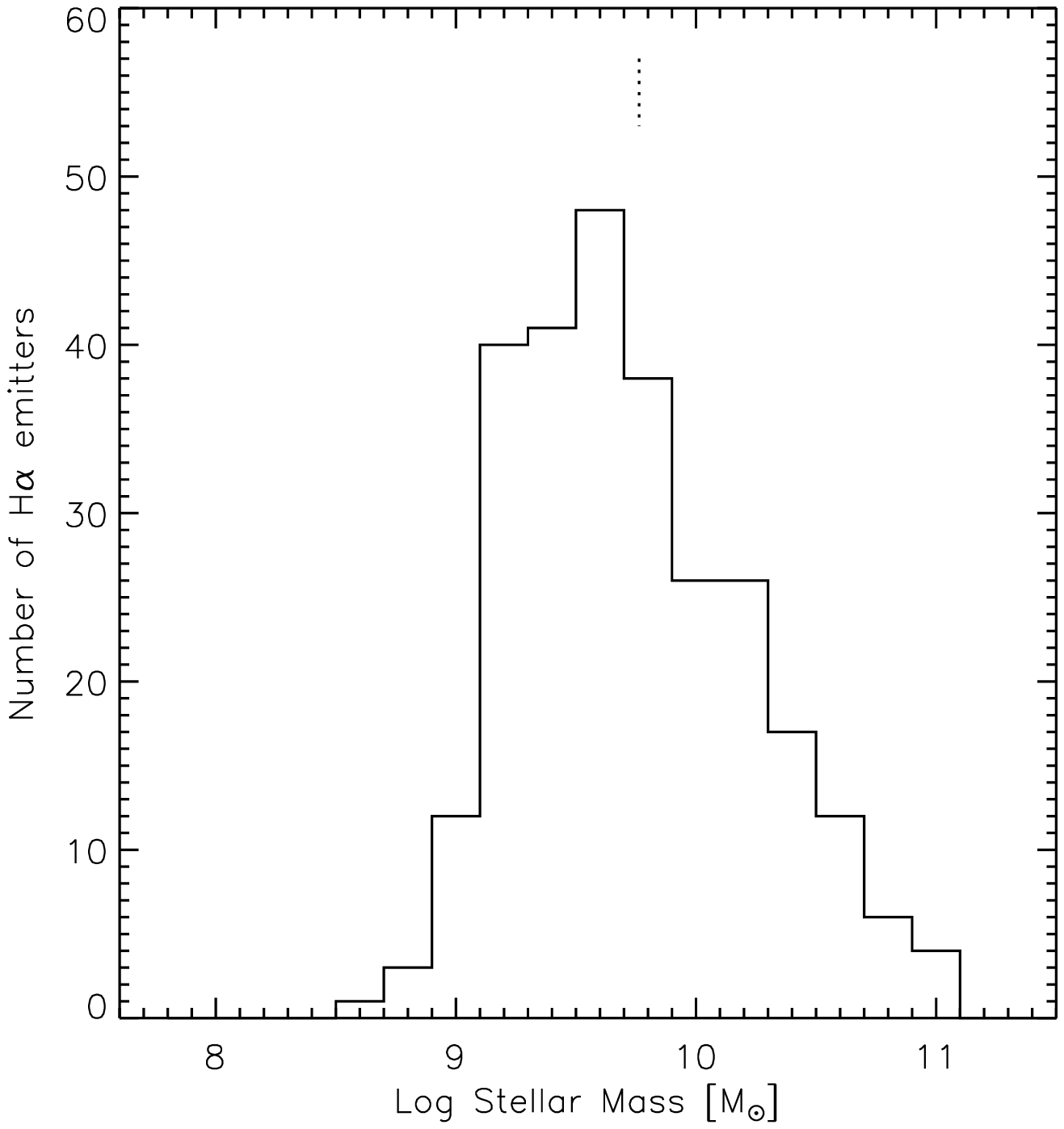}
\caption{Stellar masses for the NEWFIRM sample of $z=0.8$ \ha\ emitters in the SXDS field. The number of galaxies for which stellar masses could be derived with uncertainties of $\sim$25\% is 274.  The masses of the sample range from  $6\times10^{8} M_{\odot}$ to  $10^{11} M_{\odot}$, and the median is $6\times10^9 M_{\odot}$ (vertical dotted line).} 
\label{fig:stellarmass}
\end{figure}

The distribution of derived stellar masses for the SXDS 
spectroscopically confirmed $z=0.8$ H$\alpha$ emitters is shown 
in Figure~\ref{fig:stellarmass}.  The median of
the sample is $6\times10^9 M_{\odot}$, and 97\% have masses
between $1\times10^9 M_{\odot}$ and $4\times10^{10} M_{\odot}$.
There are a few galaxies with masses as low as $6\times10^{8} M_{\odot}$,
and as high as $10^{11} M_{\odot}$. 

\section{SDSS Comparison Sample}

To study the evolution of the dust attenuation between redshift 0.8 and today, a comparison sample of star-forming galaxies at $z\sim0$ is required.  Here, we use the data from the Sloan Digital Sky Survey, Data Release 4 \citep[SDSS DR4; ][]{sdss_dr4}.
This dataset has been reprocessed by a group from the Max Planck Institute for Astrophysics and the Johns Hopkins University (hereafter, the MPA-JHU group\footnote{\url{http://www.mpa-garching.mpg.de/SDSS/DR4/}}). The analysis pipelines for previous data releases are described in \citet{brinchmann04} and \citet{christy}. Spectroscopic redshifts and line fluxes
are available for 406,506 objects with $0.04<z<0.2$.

To select a local sample appropriate for comparison we apply a series of cuts to remove galaxies with low-quality spectra and uncertain flux measurements.  Removal of AGN from the sample is discussed in the next section. 
Sources which have uncertain redshift estimates (Z.WARNING$>0$) and redshifts at $0.146<z<0.148$ are excluded.  The latter criterion is applied because of contamination of the redshifted \hb\ line from the [OI]$\lambda$5577 sky emission line.  Multiple entries of the same object are also excluded using the list of repeats provided by the MPA-JHU group.  
Balmer line ratios are used to derive the nebular attenuations of the SDSS galaxies, so we also require that \ha\ and \hb\ are both detected at a S/N$\geq5$\footnote{The formal line flux uncertainties in the \ha, \hb, [OIII]$\lambda$5007 and [NII]$\lambda$6584 lines are increased by factors of 2.473, 1.882, 1.566 and 2.039, respectively, as recommended on the SDSS data server.}.
150,942 objects remain after these selection criteria are applied.  If the S/N$\geq5$ restriction is also placed on the \hg\ and \hd\ fluxes, the sample is reduced to 123,950 and 67,261 objects, respectively.

To compare the relationship between dust attenuation and stellar mass at $z\sim0$ and $z=0.8$, the 
stellar masses for the SDSS galaxies derived by \citet{gallazzi05, gallazi06} are used.  These masses are computed using a Bayesian likelihood estimate based on fits of the D4000, \hb, $H\delta_{A}+H\gamma_{A}$, $[Mg_{2}Fe]$ and $[MgFe]'$ measurements. The models account for dust attenuation, including negative attenuation due to errors in the index and magnitude measurements.  The stellar masses derived via this method for the SDSS sample are in good agreement with those of \citet{kauffmann03}.

To compare the attenuations of the SDSS galaxies and the NEWFIRM H$\alpha$ emitters as a function of the observed J-band magnitude for galaxies at $z=0.8$,
we shift the SDSS SEDs to $z=0.8$
and synthesize J-band magnitudes using the bandpass of the NEWFIRM J-band filter.
The SDSS SEDs are constructed using $ugriz$ "Cmodel" total magnitudes\footnote{\url{http://www.sdss.org/dr3/algorithms/photometry.html\#mag\_model}},\footnote{\url{http://www.mpa-garching.mpg.de/SDSS/DR4/SDSS\_info.html}} and the IDL \textit{kcorrect}\footnote{\url{http://howdy.physics.nyu.edu/index.php/Kcorrect}} procedures \citep[v4\_2,][]{kcorrect}, which fits restricted SED models to the photometry, using BC03 stellar evolution templates.

Finally, we also examine the dust attenuation as a function of the SFR.  To calculate the SFRs for the SDSS sample, we use H$\alpha$ fluxes measured from the SDSS spectroscopy. The \ha\ SFRs are not dust-corrected for consistency with the $z=0.8$ sample.
The SDSS \ha\ fluxes are measured within a 3\arcsec\ fiber which only samples a portion of the galaxy. Therefore we 
adopt an aperture corrections based on the difference between the fiber and Petrossian $r$\arcmin-band magnitudes as suggested by \citet{hopkins03}:
\begin{equation}
A=10^{-0.4(r'_{Petro}-r'_{fiber})}
\end{equation}
Of course, it must be kept in mind that the 
presence of radial gradients within galaxies or other spatial variations in the SFR may lead to systematic uncertainties when this correction is applied.  
For example, \citet{brinchmann04} have suggested that this aperture correction may be problematic for galaxies with stellar masses $\log M_{\ast}>10.5$ because they have large bulges which contain little star-formation. 
Discussions of the potential systematics in the estimated total SFR resulting from such corrections can be found in \citet{kochanek00}, \citet{baldry02}, \citep{gomez03}, \citet{nakamura03}, and \citet{pg03}.  
Using these corrected H$\alpha$ fluxes, the observed SFR is calculated using the \citet{kennicutt98} prescription.

\section{AGN Identification}

The objective of this study is to extend Balmer decrement measurements of the nebular attenuation of typical star-forming galaxies to intermediate redshift.  To do this we must first identify and exclude galaxies with strong AGN signatures from the samples.  Emission that is dominated by AGN light may only represent the small central region of a galaxy and not the star forming regions in general.  Furthermore, the physical conditions of the gas in the central regions of AGN-powered galaxies may result in differences in the dust creation and destruction processes from those operating in star-forming regions. 

For the purposes of our work we are only interested in removing the AGN with rest-frame optical signatures which dictates the choice of methods we use.  Our methods can be broadly divided in two categories: ones that identify Type 1 AGN 
and ones that identify Type 2/LINERs. Type 1 AGN do not obey the diagnostic developed to identify Type 2 AGN. Thus we first exclude the Type 1 and then proceed with with Type 2 identification.

We consider two methods to identify Type 1 AGN: NUV variability and optical line width. Type 1 AGN typically exhibit broad emission lines which are thought to originate from material close to the central black hole. The lines are gravitationally broadened.  In addition, the flux emitted by quasars and AGN has been found to vary aperiodically, on time scales from hours to decades at all wavelengths.  Such variability was first reported in connection with reverberation mapping \citep{peterson93,paltani94,kaspi00,kaspi07}. Evidence suggests that the majority of UV-optical variability is driven by changes in the disc accretion rates which cause variation in the UV continuum \citep{nuvvar}.  

In the SDSS sample Type 2 AGN are identified using the BPT diagram \citep{bpt}. In the NewH$\alpha$ sample, we use the Mex diagram \citep{weiner,juneau}.

In the local Universe, a method widely used to distinguish star-formation from Type 2 Seyfert and LINER-powered emission is the BPT diagnostic, which is based on the [OIII]$\lambda$5007/\hb\ and [NII]$\lambda$6584/\ha\ line ratios. \citet{kauffmann03} and \citet{kewley01} have defined criteria in this parameter space to separate galaxies by the dominant source of emission - star-formation, AGN (Seyfert 1 or LINER) or composite SF+AGN.  We apply this diagnostic method to our comparison sample of SDSS emission-line galaxies.  

At $z=0.8$, however, the \ha\ and [NII]$\lambda$6584 lines are redshifted to the NIR and are no longer accessible with optical spectroscopy, and thus the BPT diagnostic cannot be applied.  In its place we use a combination of alternative methods to identify AGN based on the overall spectral characteristics, blue spectral features, and the physical properties of the galaxies.   In the future, it will be valuable to obtain NIR spectra for the $z=0.8$ \ha\ emitters to identify AGN with the BPT diagnostic. 

The MEx diagnostic was proposed to distinguish star-formation and AGN emission at $z>0.4$ \citep{weiner,juneau}.  The method relies on the combination of [OIII]$\lambda5007$/\hb\ line ratio and the stellar mass.  The stellar mass is used as a substitute for the traditional [NII]$\lambda$6584/\ha\ ratio because it is a metallicity indicator for star-forming galaxies, and the empirical mass-metalicity relation suggests a physical connection between stellar mass and that line ratio \citep{christy}.  AGN have higher [NII]$\lambda$6584/\ha\ ratios because of the excitation conditions of the gas, but they also reside in galaxies with the highest stellar masses \citep{kauffmann03}. An alternative diagnostic diagram which uses the $(U-B)$ color instead of stellar mass is presented by \citet{yan11}. This color-excitation diagnostic agrees well with the MEx method thus here we have decided to only use the latter. 


The objects identified as AGN with the various methods in this Section are listed in Table~\ref{tab:agn}.

{\bf NUV variability: }The GALEX data, which include six epochs of observations spaced by $\sim$year-long timescales, allow us to look for variations in the rest-frame FUV $983-1572$\AA\ (observed NUV) emission.  One variable \ha\ emitter, whose observed NUV flux varies by 1.2 magnitudes, is identified.  The optical spectrum of this object is dominated by broad emission lines. 

{\bf Broad Lines: }The routine used to fit the continua in the IMACS spectra also provides coarse spectral type classifications for the objects based on the best fitting template.   Of all 341 \ha\ emitters, five are identified as QSOs and/or broad-lined AGN.   
The NUV variable discussed above is also identified as an AGN by the fitting routine.
The continua of two objects identified as hosting broad-lined AGN are not well-fit and are excluded from the remainder of the analysis.  Finally, all of the spectra for the $z=0.8$ H$\alpha$ emitters were also visually inspected, and two additional objects with broadened lines which were not identified by the routine are found. 

{\bf The MEX diagnostic: } To apply the MEx diagnostic to our sample, we require that both the [OIII]$\lambda5007$ and the \hb\ lines are detected with S/N$\geq3$. Of the 341 emitters, 221 fit this requirement. As noted above, a fraction of the sample objects do not have mass estimates due to lack of optical photometry (the 42 emitters in SA22 and 12 of the emitters in SXDS). The objects with poor SED fits (13 objects in SXDS) are also excluded. In the end, the method can be applied to 175 of the 341 $z=0.8$ H$\alpha$ emitters with IMACS spectra.  

\begin{figure}
\figurenum{6}
\label{agn}
\epsscale{1.2}
\plotone{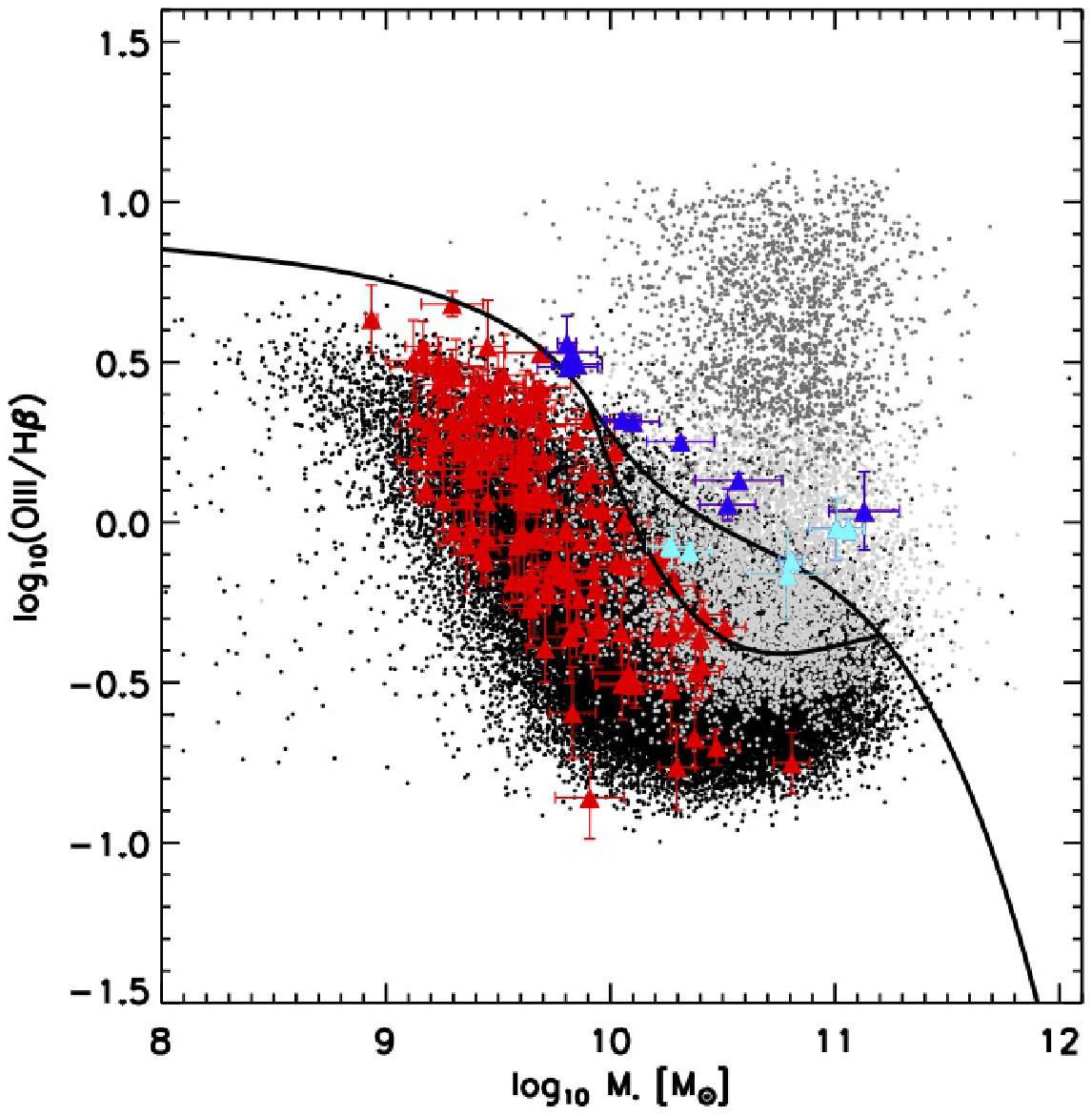}
\caption{The Mex diagram  \citep{juneau} is used to quantify the AGN contamination of the sample. The local SDSS sample is split by their position on the traditional BPT diagram: star-forming (black), composite (light gray) and AGN (dark gray). Triangles show the objects from the New\ha\ sample with $1\sigma$ errors. We require that the [OIII]$\lambda5007$ and \hb\ line be detected with S/N$\geq3$. The MEx diagram identifies nine AGN candidates (blue points) out of a sample of 180 galaxies for which we have both line measurements and a mass estimate. We also identify seven candidate AGN-SF composite objects (cyan triangles). The remaining objects are normal star forming galaxies (red points). }
\end{figure}

Figure \ref{agn} shows the MEx diagnostic for the $z=0.8$ H$\alpha$ emitters and the SDSS DR4 comparison sample. There is an excellent overlap between the loci of the two samples. In addition to this general demarcation of the regions predominantly occupied by the different types of objects, \citet{juneau} have also made public their code which, given line fluxes and mass measurements with errors, calculates the probability that a given object belongs to one of four groups: star-forming galaxy, composite object, LINER or Type 2 Seyfert galaxy. We use these probabilities to identify the emission source for the objects in our sample (after discarding objects where the emission lines are affected by poor sky line subtraction). We identify nine AGN and seven composite objects (Figure \ref{agn}). Therefore the AGN fraction in the New\ha\ sample is at least 5\% and at least  4\% of the objects are composite. There are seven objects which lie close to the line separating the star-forming and composite galaxies and have almost equal probabilities to be part of the two classes. In these cases we select the highest probability, which is for a star-forming galaxy for five of them. 

\begin{figure*}
\figurenum{7}
\label{fig:sa22mass}
\epsscale{1.1}
\plotone{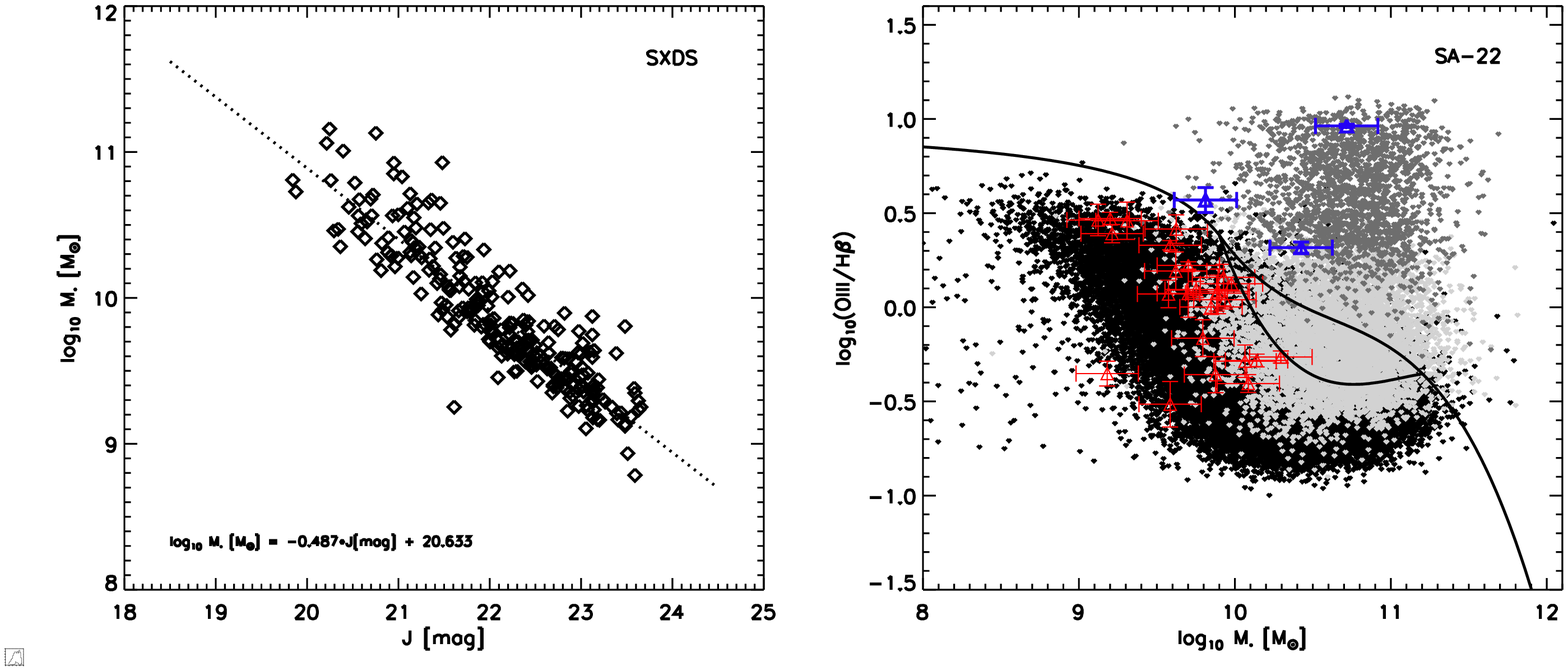}
\caption{{\it Left: } The observed J-band magnitude correlates tightly with the stellar mass of the galaxy for the \ha\ emitters in SXDS sample. This relation allows us to determine the masses of the SA-22 \ha\ emitters, for which PSF-matched broad band photometry has not yet been produced. {\it Right: } The MEx diagnostic diagram for the SA-22 sample, using the J magnitude - stellar mass relation allows us to identify three AGN (blue triangles). The local SDSS sample is also presented for comparison with symbols identical to those in Figure \ref{agn}.}
\end{figure*}

Stellar masses are not available for the SA-22 sample. We use the tight correlation between J magnitude and stellar mass to place the SA-22 \ha\ emitters on the MEx diagram as shown in Figure \ref{fig:sa22mass}. Once the linear relation is removed, the residual scatter in the mass is 0.2 dex which we use as an error on the mass estimate. In this way we identify three AGN candidates in this field.

The MEx diagnostic is defined at $z=0$ and we might expect that it will evolve with redshift due to the evolution of the physical properties of galaxies. Such evolution may be caused by the evolving mass-metallicity relation \citep[e.g.,][]{erb06}, downsizing, additional AGN contribution \citep[e.g,][]{groves06, wright10} or changing physical conditions in the HII regions \citep[e.g.,][]{brinchmann08,hainline09}. However, further studies of intermediate redshift galaxies are required to quantify these evolutionary effects and disentangle them from the possible selection effects of current samples. \citet{juneau} find that such effects are likely minor because at $0.3<z<1$ the MEx-selected AGN have X-ray properties consistent with those of AGN -- they are either X-ray luminous or have a hard X-ray spectrum when stacked. Star-forming MEx selected galaxies
show only a soft signal in X-ray stacks, consistent with purely star-forming systems.


Combining the results from the different methods yields 17 AGN candidates and seven composite galaxies (Table~\ref{tab:agn}).
It is unlikely that we have identified all AGN in the sample, so this is a strict lower limit on the number of AGN in the spectroscopically confirmed $z=0.8$ H$\alpha$-selected sample. We are unable to apply the MEx diagnostic to 47\% of our emitter sample because we either have no mass measurements or the S/N in the lines is lower than our limit.  However, we are likely to have identified the strongest, unobscured AGN and these are the ones which might affect the Balmer line ratios.
If the 17 objects identified as AGN and the seven objects identified as composite are the only ones in the sample, then the AGN fraction is at least 5\% and the composite fraction is at least 2\%.  This AGN fraction is consistent with the findings of \citet{garn09} who also examine the AGN contamination  in a sample  of $z=0.84$ \ha-selected emitters similar to ours. They identify $5-15\%$ of their sample as AGN using a broader range of methods which includes NIR color-color diagnostics, mid-to-far IR slope and X-ray emission. Again, we underscore that for the purposes of this work, we are only concerned with removing AGN with rest-frame optical spectral signatures, which explains why our fraction is at the lower end of their range. 

We remove AGN from the SDSS sample using the classical BPT diagram \citep{bpt}. To apply the diagnostic, we require S/N$\geq3$  for [OIII]$\lambda$5007 and [NII]$\lambda$6584 as a minimum requirement for these lines to be reliably detected, in addition to the S/N$\geq5$ cut imposed on the Balmer lines in Section 4. The cuts on [OIII]$\lambda$5007 and [NII]$\lambda$6584 remove only a total of 2,998 objects, for a final sample of 64,263 objects. The most stringent requirement on the sample is the \hd\ line detection, thus the sample is effectively \hd-selected.  Using the \citet{kewley01} and \citet{kauffmann03} relations, 6,629 objects (10.3\% of the sample) are identified as composite, 2,417 objects (3.8 \% of the sample) are identified as AGN and the remaining 55,217 (85.9 \%) -- as star-forming. 
In Figure \ref{agn} these three classes are plotted in the parameter space of  the MEx diagnostic  \citep{juneau}. In order to be able to apply this diagnostic we also require that [OII]$\lambda$3727 has S/N$>3$ (removing 116 objects) and that the objects have mass measurements  (removing 17,565 galaxies). 

\begin{figure*}
\figurenum{8}
\label{sdss_comp}
\epsscale{1.1}
\plotone{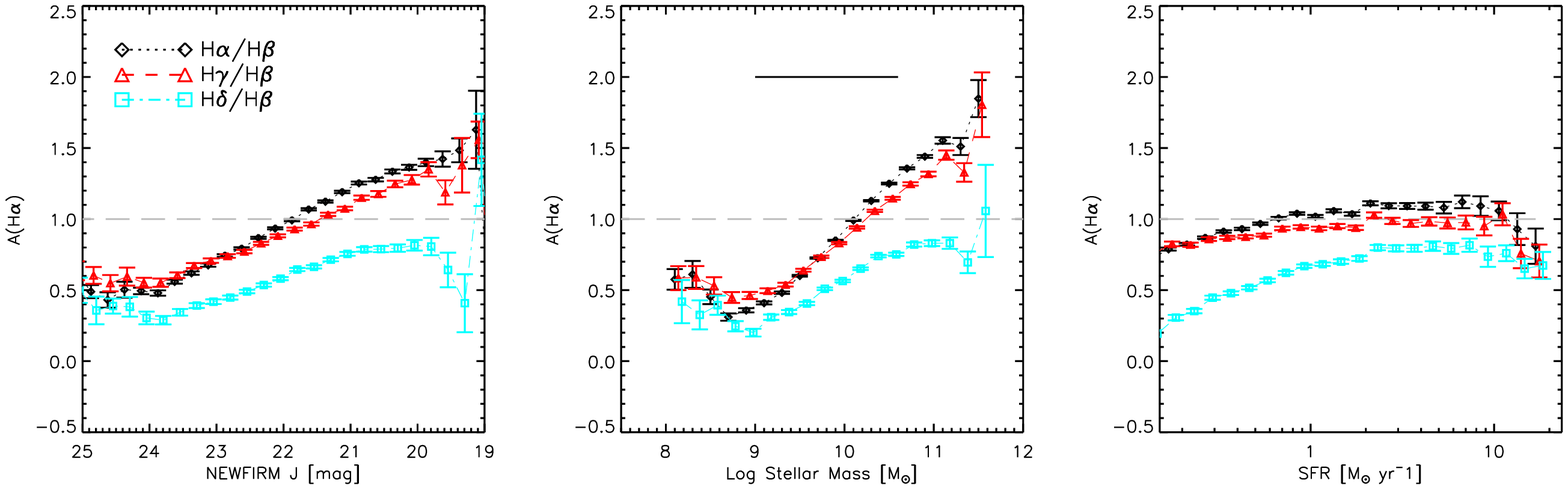}
\caption{Comparison between the attenuation A(\ha) derived from different Balmer line ratios in the SDSS sample as a function of J magnitude, stellar mass and SFR. Shown are the binned weighted means (open symbols, lines) and the weighted standard deviation in each bin. The bins are slightly offset for clarity. The dashed grey horizontal line indicates A(\ha)=1 for reference. In the middle panel the range of masses of the New\ha\ sample  is also indicated (vertical solid line). The \ha/\hb\ and \hg/\hb\ ratios yield consistent attenuation as a function of all parameters. The \hd/\hb\ ratio deviates at bright magnitudes, high masses and low star-formation rates \citep[see ][]{balmersdss}.}
\end{figure*}

\section{Balmer Decrement Attenuation Measurements}

Comparison of the observed ratio between any two Balmer lines to the expected recombination ratio (the Balmer decrement) yields a measure of the nebular dust attenuation, given an extinction curve. 
A detailed description of how the attenuation is calculated from the observed Balmer ratios is given in the Appendix. Here, we assume standard Case B recombination with temperature $T=10^{4}$ K and electron density $n_{e}=10^{2}$ cm$^{-3}$, which results in expected ratios of 2.86, 0.47 and 0.26 for H$\alpha$/H$\beta$, H$\gamma$/H$\beta$ and H$\delta$/H$\beta$, respectively \citep{osterbrock89}.
We also assume the \citet{calzetti97} attenuation law to convert the color excess to attenuation. 

The present paper is focused on the high order Balmer ratios \hg/\hb\ and \hd/\hb. Of course, the $z=0.8$ sample is \ha-selected, and we do have measurements of the \ha\ line fluxes. However, we do not include analysis of the \ha\ Balmer decrements in this work because it requires a detailed  examination of the possible sources of systematic error when combining the photometric and spectroscopic data and it is beyond the scope of the current paper. The challenges in measuring the \ha/\hb\ ratio with the existing data are discussed in more detail in Section 7.2.

\subsection{Nebular Dust Attenuation in a Comparison Sample of Local Galaxies}

Before examining the results for the $z=0.8$ H$\alpha$-selected galaxies, we review the attenuation properties of the local SDSS sample. The H$\alpha$ attenuation, $A(H\alpha)$, is computed from the \ha/\hb\ line ratio, as well as from \hg/\hb\ and \hd/\hb.  In Figure \ref{sdss_comp}, the mean weighted A(H$\alpha$) is shown as a function of J magnitude, stelar mass and SFR.  The attenuation measurements derived from \ha/\hb\  and from \hg/\hb\  are consistent and the trends are in agreement. However, the attenuation derived from the \hd/\hb\ ratio shows systematic offsets. The attenuation based on \hd/\hb\ is lower at bright magnitudes, high masses and low star-formation rates. Such an offset is also
 reported by \citet{balmersdss} based on measurements from the SDSS DR7 dataset and is thought to be caused by a systematic error in the continuum fit around the \hd\ line.   The exact source of the discrepancy, however, is still being investigated by the SDSS collaboration (J. Brinchmann, private communication).  We do not try to correct for this offset in our analysis, and in the figures that follow, we will compare our $z=0.8$ sample to the SDSS \hg/\hb\ sample, unless otherwise noted.  The median attenuations and standard deviations from the three line ratios are $0.98\pm0.43$, $0.97\pm0.82$ and $0.62\pm0.90$, respectively.  
 
 The scatter dramatically increases when A(H$\alpha$) is based on the \hg/\hb\ and \hd/\hb\ ratios because of the weakness of the lines, the limited wavelength range spanned by them and the associated increase in the measurement errors.  The large scatter is one reason for the larger fraction of objects exhibiting negative attenuation: while only 0.5\% of galaxies have negative attenuation when A(H$\alpha$) is computed from \ha/\hb\, the fraction increases to 8\% and 21\% when \hg/\hb\ and \hd/\hb\ are used.  

\subsection{Comparison between Balmer Decrements}

\begin{figure}
\figurenum{9}
\label{comparison}
\epsscale{1.0}
\plotone{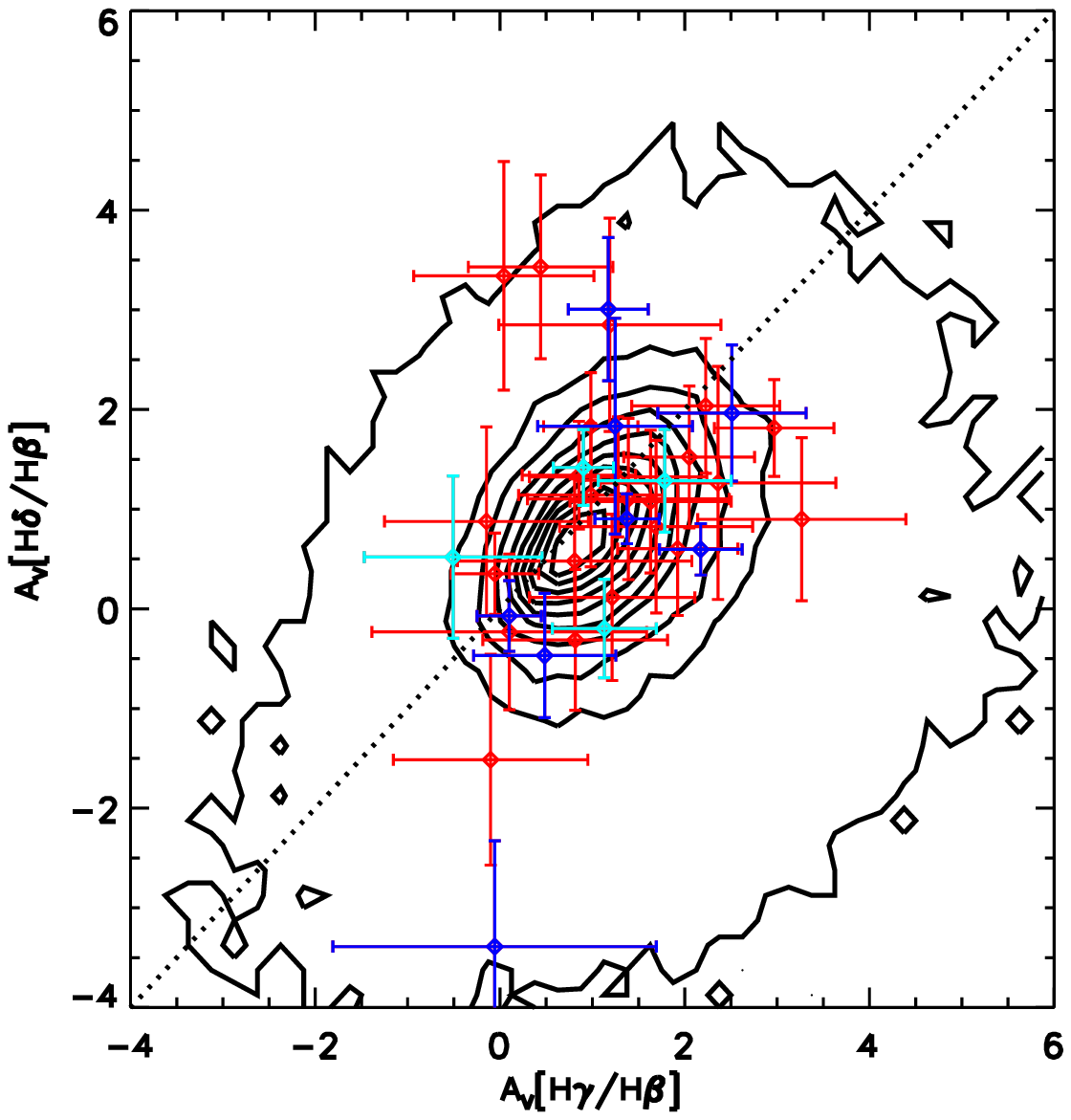}
\caption{Comparison between the values of the broadband attenuation $A_{v}$ as derived from $H\gamma/H\beta$ and from $H\delta/H\beta$ for a subsample of 34 objects where all three lines are detected at S/N$>5$. Contours mark the locus of SDSS galaxies. The one-to-one line is shown (dotted line). The correlation between the two values is significant at the 95\% level.}
\end{figure}

For a sample of 35 objects we have measurements of both the $H\gamma/H\beta$ and the $H\delta/H\beta$ ratios and can compare the attenuations derived using the two Balmer decrements (Figure \ref{comparison}). The scatter between the two measurements for the same objects is significant. However, the mean values of the attenuation inferred separately from the \hg/\hb\ and \hd/\hb\ are  consistent with the one-to-one relation, within the 1$\sigma$ measurement uncertainties (Figure \ref{comparison}). The Pearson correlation coefficient for the sample is 0.33 and the Spearman correlation coefficient is 0.38. These correspond to ~95\% significance of the correlation. The mean (median) difference between the two measurements is 0.19 (0.33) $\pm1.3$ mag  (1 $\sigma$), generally consistent with no systematic. 

\subsection{Balmer Decrements in $z=0.8$ H$\alpha$-selected Galaxies}

The depth of our IMACS spectroscopy enables the detection of \hb\ and \hg\ lines with S/N sufficient for the determination of the nebular attenuations for 16\% of our spectroscopically confirmed \ha\ emitters at $z=0.8$ (N=55).  The \hb\ and \hd\ lines are detected with sufficient S/N for the analysis for 15\% of the sample (N=50).  In this section, we first examine the distribution of the observed Balmer line ratios of the $z=0.8$ H$\alpha$ emitters and the SDSS sample, and then compare the distributions of A(H$\alpha$) for the $z=0.8$ and the SDSS samples.

\begin{figure*}
\figurenum{10}
\label{balmer1}
\plotone{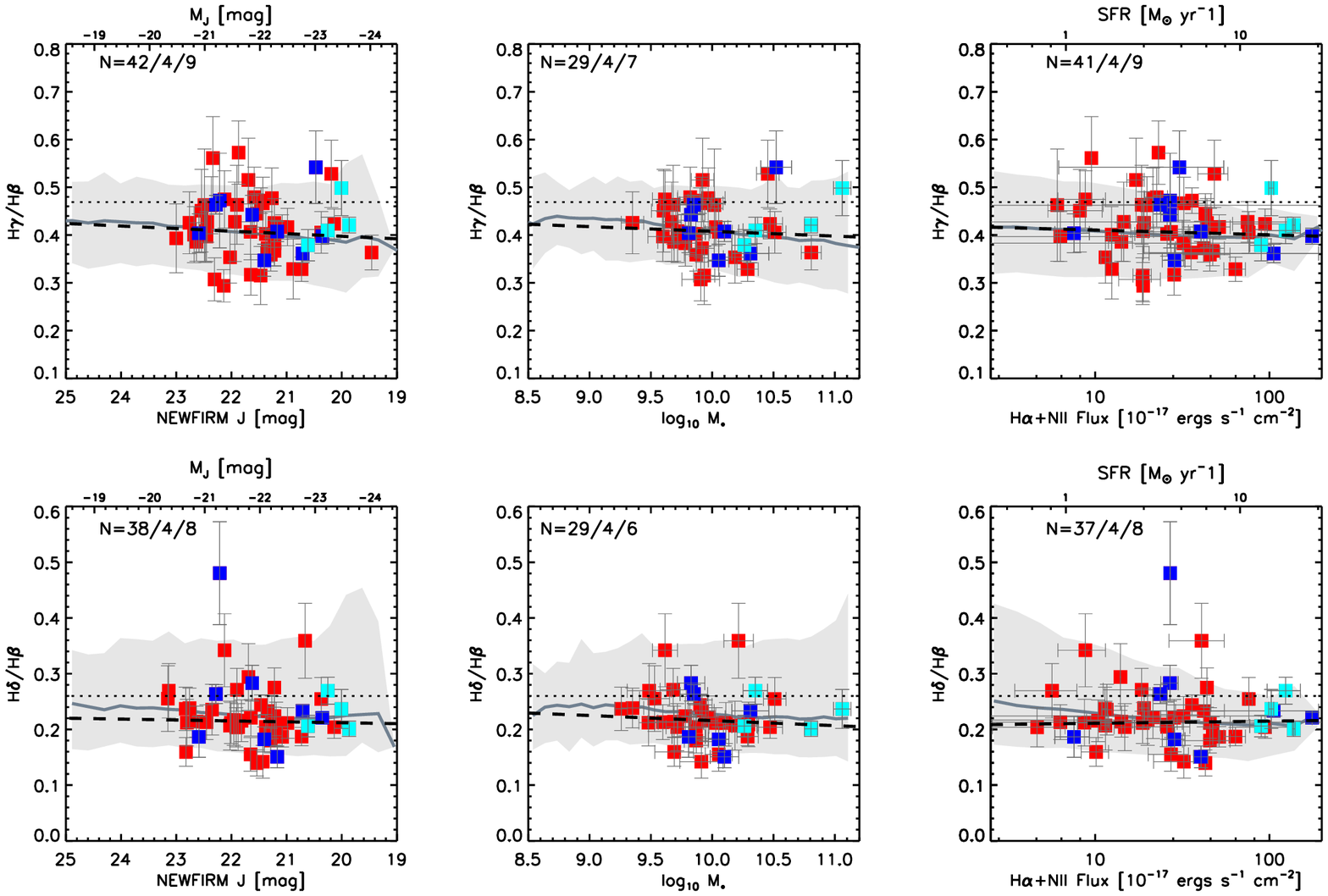}
\caption{Observed Balmer line ratios for individual $z=0.8$ galaxies as a function of J-band magnitude (left column), stellar mass (middle column), and $H\alpha+$[NII]$\lambda$6584 flux (right column).  Objects with \hg/\hb\ line measurements are shown on the top row and those with \hd/\hb\ line measurements on the bottom row. The numbers of star-forming galaxies (red symbols), galaxies with composite emission (cyan symbols) and galaxies hosting AGN (blue symbols) included in the analysis are indicated in the top left corners of the panels.  The dotted lines indicate the ratios expected under the conditions of Case B recombination and the dashed lines indicate the least squares fit to the individual line ratios. In all cases, the slope of the least squares fit is in the expected direction - that the line ratios decrease with decreasing magnitude, increasing stellar mass and increasing $H\alpha+$[NII]$\lambda$6584 flux. Also shown is the conditional distribution for the local SDSS sample (0.05 to 0.95 percentile, shaded light gray). The local and $z=0.8$ samples occupy similar loci and the least squares fits (black, dashed line) agree well with the 0.5 percentile contour (dark gray, solid line). The offset of the local \hd/\hb\ sample towards higher line ratios is due to a systematic error and discussed in Section 6.1 and \citet{balmersdss}.}
\end{figure*}

The observed Balmer line ratios for individual $z=0.8$ galaxies as a function of J-band magnitude, mass and \ha+[NII] flux are shown in Figure \ref{balmer1}.  The \hg/\hb\ line ratios are plotted in the top row, and the \hd/\hb\ line ratios in the bottom row.  The difference in the number of objects plotted in each panel is due to variations in the available data.  The \hd/\hb\ sample is not entirely a subset of the \hg/\hb\ sample.  In some cases the \hg\ line is contaminated with sky emission but the \hd\ line is clear. Such objects are included in the \hd/\hb\ sample but excluded from the \hg/\hb\ sample.   

Figure 10 shows a fair amount of scatter in the observed line ratios. First order linear fits to the data indicate relations in the expected directions: that the ratios decrease (i.e. attenuation increases) towards brighter magnitudes, higher masses and higher SFR. However the data are equally consistent with a constant ratio as they are with a linear fit (based on the reduced $\chi^{2}$ of the fits). The mean (median) line ratios in the total sample as a function of J-band magnitude (the largest sample) are \hg/\hb$=0.42 (0.42) \pm0.06$  and  \hd/\hb$=0.23 (0.22) \pm0.06$. When the AGN and composite galaxies are excluded, the mean line ratios and  standard deviations are: $0.42\pm0.07$ and $0.22\pm0.04$, respectively.  Figure \ref{balmer1} also compares the $z=0.8$ sample with the conditional distribution of the $z=0$ SDSS sample (i.e., each bin in the distribution is normalized separately). For the \hg/\hb\ sample, the distribution of the $z=0.8$ line ratios is consistent with that of the $z=0$ sample and the least squares fit to the $z=0.8$ data agrees well with the 0.5 percentile local contour. The systematic shift in the SDSS \hd/\hb\ values towards higher lnie ratio (i.e., lower attenuation) as discussed in Section 6.1 creates an apparent difference between the SDSS weighted mean and the linear fit to the $z=0.8$ data.  The mean (median) line ratios (AGN included) correspond to an attenuation for the New\ha\ sample of $0.8(0.9)\pm1.0$ (\hg/\hb), $0.7(0.9)\pm1.0$ (\hd/\hb) and $0.8(0.9)\pm1.0$ (joint) for the J-magnitude sample,  which are consistent with the average attenuation from the \hg/\hb\ line ratio in the SDSS sample ($1.0\pm0.8$).  

Some of the New\ha\ objects exhibit line ratios above the Case B line which would indicate negative attenuation. Rather than having unphysical line ratios, these objects are likely scattered up due to errors in their line measurements. The offsets above Case B are generally consistent with the measurement errors. The local SDSS sample also exhibits a spread of points which populate the area of the plot above Case B.

\begin{figure*}
\figurenum{11}
\label{balmer2}
\plotone{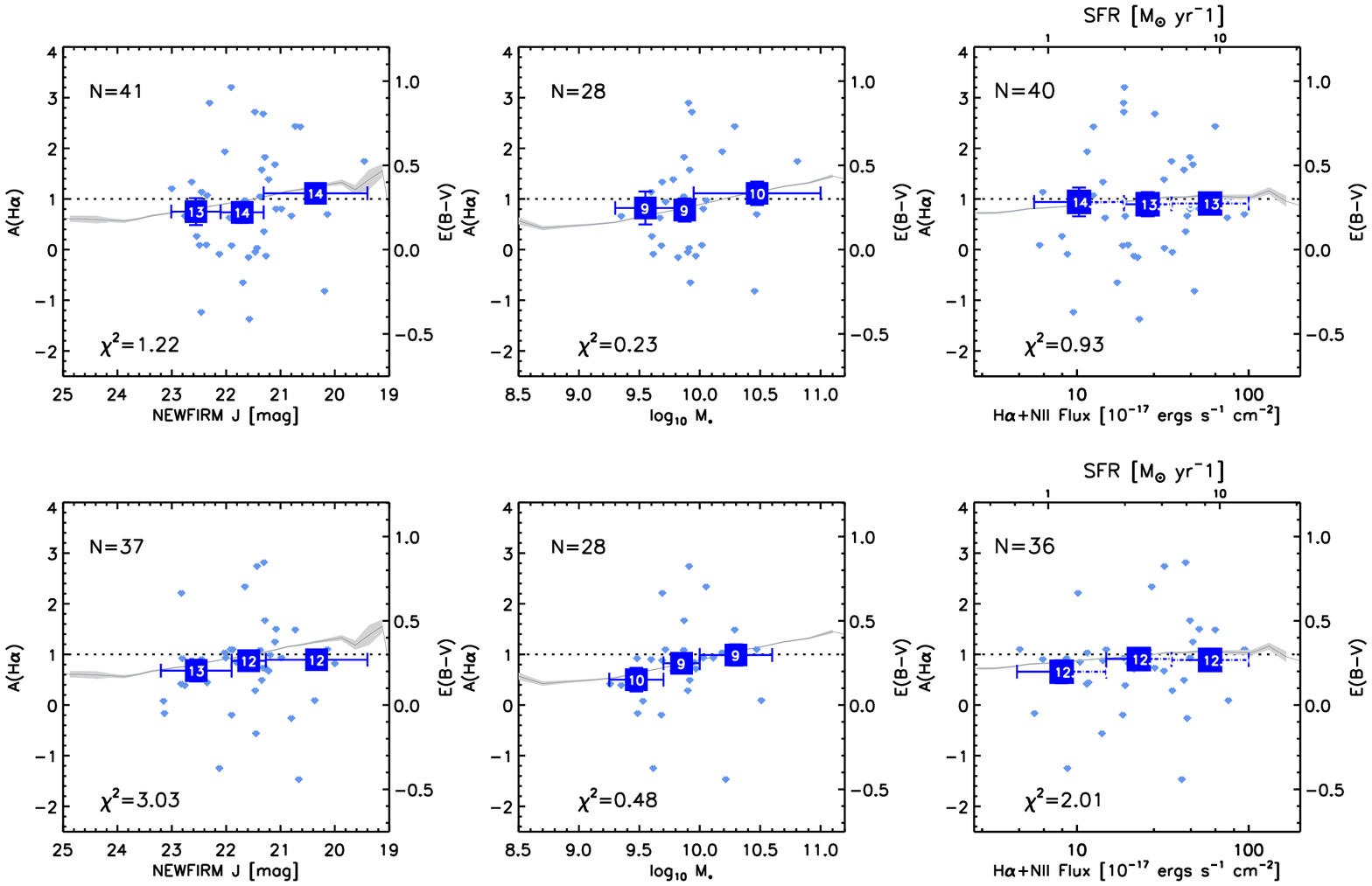}
\caption{Comparison between the weighted mean Balmer decrements of $z=0.8$ star-forming galaxies and of the local SDSS sample contours as a function of J-band magnitude (left column), stellar mass (middle column), and $H\alpha+$[NII]$\lambda$6584 flux (right column). Blue squares indicate the binned weighted mean and the weighted error of the mean for the $z=0.8$ sample with the number of galaxies in each bin indicated. The individual measurements are also shown (small light blue symbols). For the local sample we also indicate the weighted mean and the standard error of the mean (dark gray line, light gray shading). The fiducial value of A(\ha) $=1$ magnitude is also shown (dotted line). We display separately the attenuation derived from the \hg/\hb\ (top row) and from the \hd/\hb\ (bottom row) decrements. The $z=0.8$ sample exhibits trends consistent with those of the $z=0$ sample as indicated from the $\chi^{2}$ estimates in each panel.}
\end{figure*}

The overall trends of A(H$\alpha$) based on the individual line measurements for the $z=0.8$ and the $z=0$ samples are shown in Figure~\ref{balmer2}, again as a function of J-band magnitude, stellar mass and \ha+[NII] flux. We show the binned weighted mean and weighted errors on the mean for the two samples where the measurement errors on the individual points are used as weights\footnote{We use the routine {\it meanerr.pro} which is adapted by Marc Buie from "Data Reduction and Error Analysis for the Physical Sciences", p. 76, Philip R. Bevington, McGraw Hill}. The New\ha~ points are binned such that each bin contains one third of the sample (blue points). The results from the \hg/\hb\ line ratio (top row) and the \hd/\hb\ line ratio (bottom row) are both compared to the \hg/\hb-derived SDSS attenuation. Here we have excluded all New\ha\ galaxies identified as AGN and composite objects. In each panel we also show the total $\chi^{2}$ of the difference between the  New\ha~ and SDSS attenuation:
\begin{equation}
\chi^2 = \sum\frac{(A(H\alpha, SDSS) - A(H\alpha, NewH\alpha))^2}{\sigma(A(H\alpha, SDSS))^2+\sigma(A(H\alpha, NewH\alpha))^2}
\end{equation}
In all panels the $\chi^{2}$ is small indicating that the weighted means of the two samples are very similar. As in Figure~\ref{balmer1} we also compare the reduced $\chi^{2}$ of a linear fit to the New\ha~ binned points with the $\chi^{2}$ of a constant ratio. The linear fits in all panels have smaller $\chi^{2}$ than the constant ratio, indicating significant linear trends. These differences are smaller for the trend as a function of \ha+[NII] flux and larger for the trends as a function of J-magnitude and stellar mass.

\subsection{Balmer Decrements from Stacked Spectra}
 
 \begin{figure}
\figurenum{12}
\label{balmer1distr}
\epsscale{1.2}
\plotone{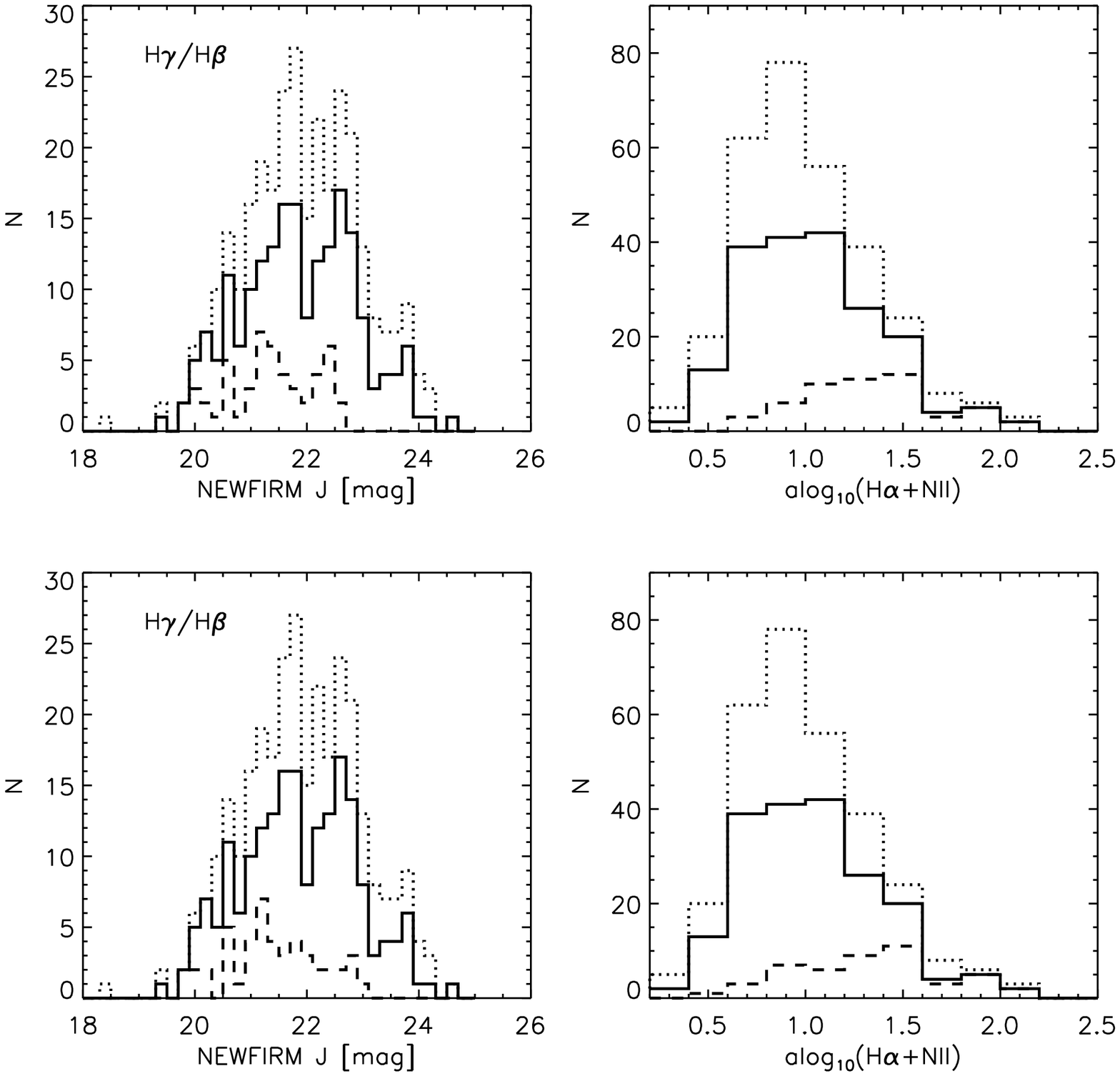}
\caption{{\it Top row:} Comparison between the full spectroscopic sample (dotted lines) and the $H\gamma/H\beta$ sub-sample (dashed lines) in terms of $J$-band magnitude (first panel) and $H\alpha+$[NII]$\lambda$6584 flux (second panel). Also shown is the sample which excludes the spectra contaminated by sky lines (solid line).
 The objects in the $H\delta/H\beta$ sub-sample have brighter $J$ magnitudes and larger $H\alpha+$[NII]$\lambda$6584 fluxes than the underlying sample. This means that the attenuation measurements based on this sub-sample will not be valid for low mass, lower-SFR galaxies, but preferentially biased to high mass, high-SFR galaxies. {\it Bottom row:} Same as the top row but for the $H\delta/H\beta$ sub-sample. The shift towards brighter $J$ magnitudes and high  $H\alpha+$[NII]$\lambda$6584 fluxes is even more evident.}
\end{figure}

 In Section 6.2 Balmer decrement measurements for individual galaxies were examined as a function of J-band magnitude, stellar mass and \ha$+$[NII] flux. However, the individual measurements exhibit a great deal of scatter, due to both intrinsic variations among the galaxies and to measurement errors. The measurement uncertainties stem from the limitations to proper continuum subtraction of the faint, high-order Balmer lines, and from measuring the line fluxes from low S/N spectra. The median error in the computed attenuation for individual galaxies in the \hg/\hb\ sample is 0.8 magnitudes, and 0.7 magnitudes in the \hd/\hb\ sample. The corresponding errors in the SDSS sample are 0.6 and 0.3 magnitudes for \hg/\hb\  and \hd/\hb, respectively. 
However, whereas the SDSS shows statistically significant correlations with the three properties plotted \citep[e.g., as discussed by ][]{garn10}, 
the errors, combined with the relatively small size of the sample of $z=0.8$ H$\alpha$ emitters with sufficient S/N measurements of the H$\beta$, H$\gamma$, and H$\delta$ lines, obscures any potential trends. Furthermore, the objects with individual Balmer line measurements are likely a biased sample and will tend to have higher luminosities, higher SFRs, higher masses and/or lower attenuation. 

\begin{figure}
\figurenum{13}
\label{fig:stack_j}
\plotone{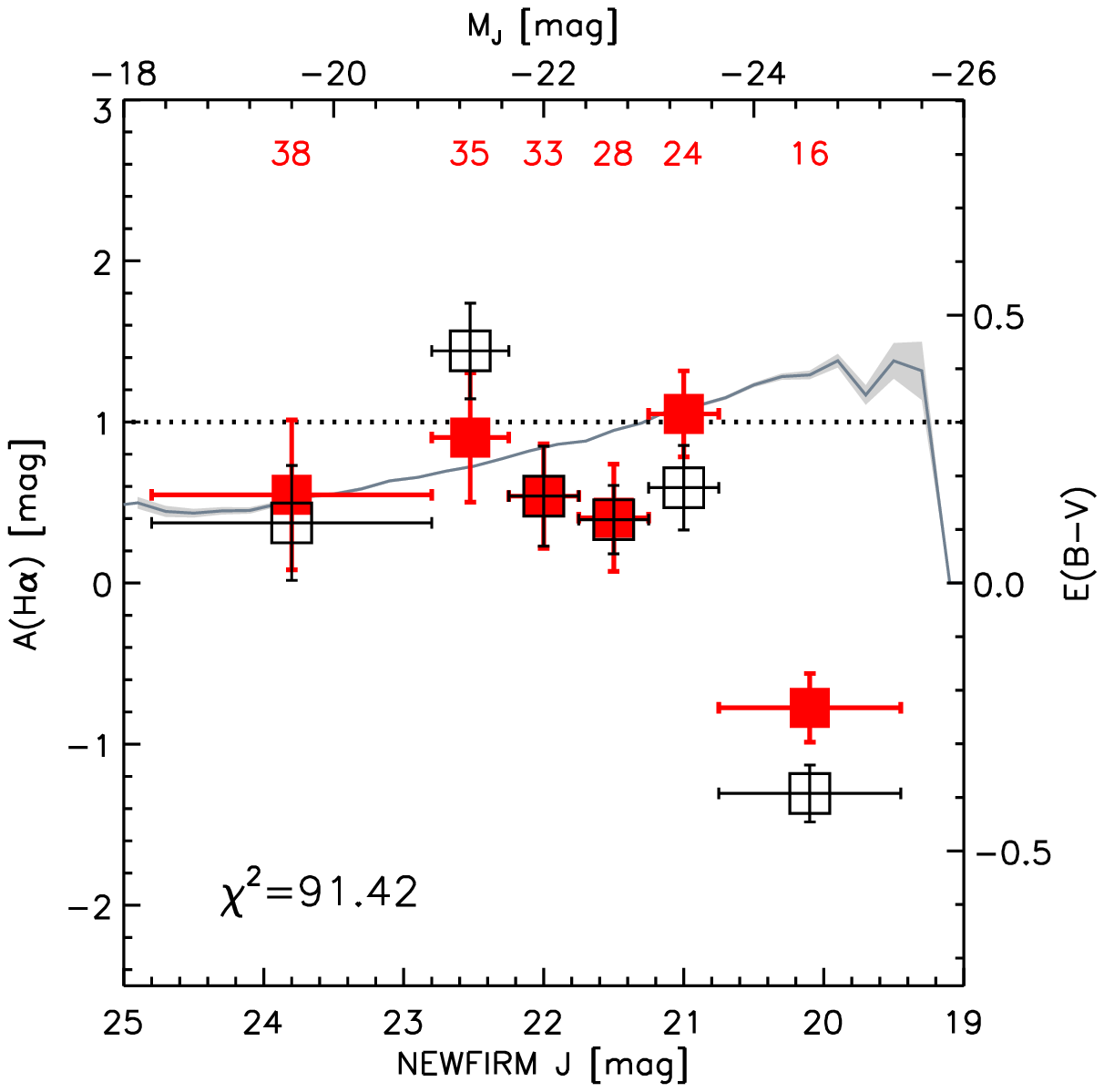}
\caption{
Mean attenuation based on the Balmer decrements measured from stacked spectra. Stacked spectra are created in bins of apparent NEWFIRM J-band magnitude. Each spectrum in the stack is weighted by the inverse variance. The top axis indicates the corresponding absolute J-band magnitude for the $z=0.8$ sample. The stacking is performed for all objects, including AGN and composite objects (open squares) and excluding both composite and AGN objects (filled red squares). All errors are based on the line measurement errors. The number of $z=0.8$ spectra in each bin is indicated in red above the symbols. Also shown is the error-weighted mean attenuation for the local SDSS sample as a function of magnitude and the $1\sigma$ weighted error on the mean (dark gray line, light gray shading).  The horizontal dotted line indicates the fiducial value of $A(H\alpha)=1$ magnitude. $\chi^2=91.42$ is a measure of the differences between the two samples. The bins and the corresponding line fluxes are listed in Table \ref{tab:data1}. Figure \ref{fig:stacked_spectra} shows the binned spectra. }
\end{figure}
 
 \begin{figure*}
\figurenum{14}
\label{fig:stacked_spectra}
\includegraphics[angle=0,width=6.5in]{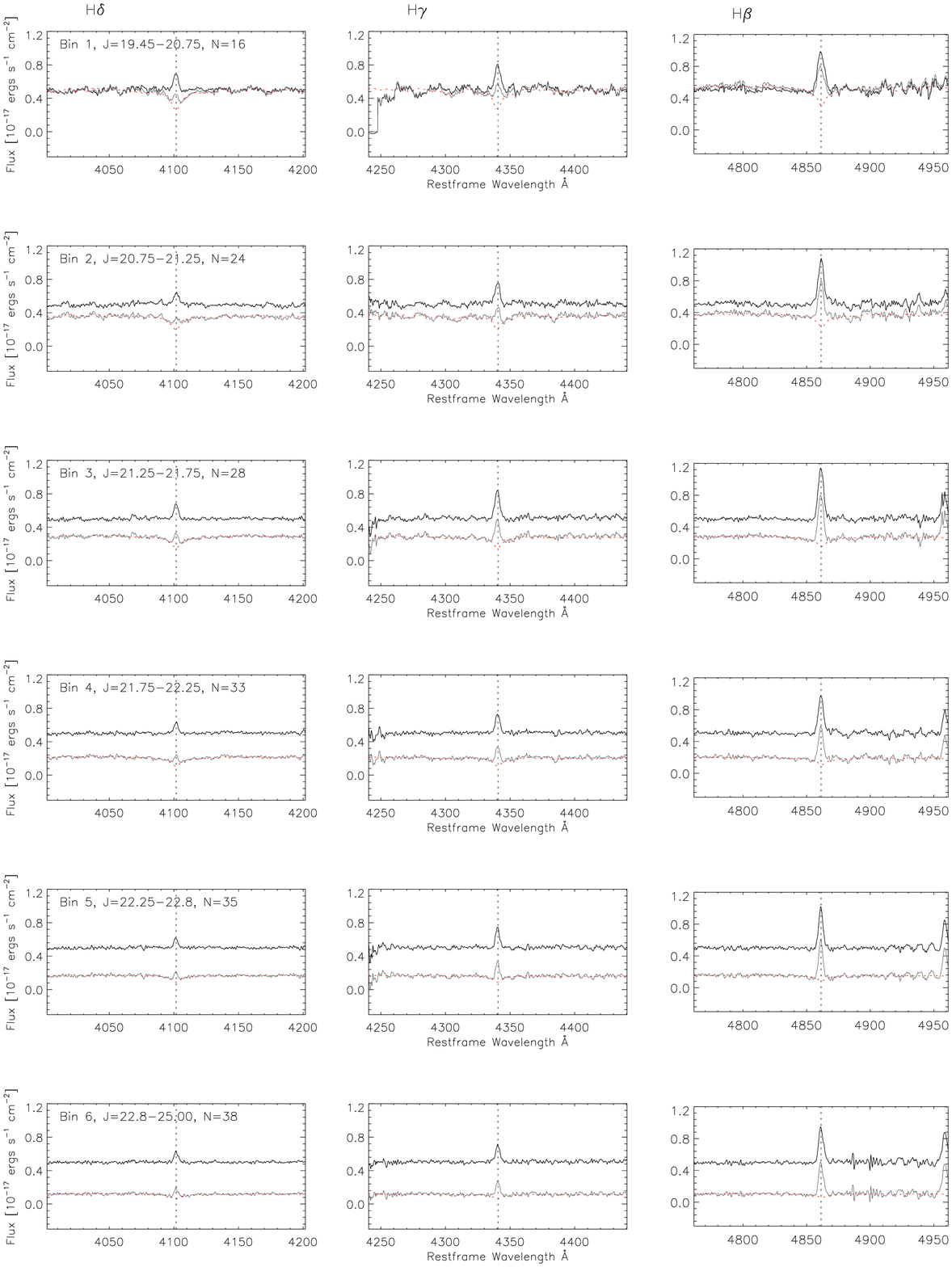}
\caption{Weighted stacked spectra, divided in six bins by J-band apparent magnitude. Shown are the stacked spectra (gray line), the continuum fits (red dotted line) and the continuum subtracted spectrum (offset by 0.5 vertically, black line). The magnitude increases top to bottom. Vertical dotted lines indicate the positions of the Balmer lines \hd\ (left), \hg\ (center) and \hb\ (right).}
\end{figure*}

 The last point is demonstrated in Figure \ref{balmer1distr}, which compares the full sample of emitters with spectroscopic follow up and the sample of emitters with individual Balmer line detections. The objects with individual detections are more luminous, having brighter J-band magnitudes. The median J-band magnitude of the sample with individual line detections ($J_{med}=21.4$) is 0.6 magnitudes brighter in J than the median for the full sample ($J_{med,full}=22.0$). Faint galaxies are underrepresented with only one object fainter than $J=23$ mag, with an individual line detection. The distribution of $H\alpha+$[NII]$\lambda$6584 fluxes is also skewed toward higher values: the median line flux for the full spectroscopic sample is 13$\times10^{-17}$ ergs s$^{-1}$ cm$^{-2}$, while for the subsample with line detection the median is 27$\times10^{-17}$ ergs s$^{-1}$ cm$^{-2}$.  {As a result of contamination from sky lines at the positions of the Balmer lines, approximately half of the sample is excluded from the stacks. However, the median J magnitude and $H\alpha+$[NII]$\lambda$6584 flux of the remaining sample are very close to those of the full sample ($J=22.0$ and $H\alpha+$[NII]$=11.8$) indicating that excluding the objects with sky line contamination does not bias the sample. Furthermore, a comparison between the distributions of the values of $\tau_{V}$ (from the SED fits) for the included and excluded samples does not reveal a bias in the properties of the included sample.}
 
In order to address these issues, we create stacked spectra in bins of J-band magnitude, stellar mass and \ha$+$[NII]$\lambda$6584 flux. The stacks contain both objects with and without individual detections of the Balmer lines. The bins are selected in a manner which achieves a minimum S/N of 10 in the \hb\ line and a minimum of 10 objects per bin. This leads to uneven number of objects per bin (fewer objects in brighter bins) and uneven bins. The stacked spectra are created by co-adding the inverse-variance-weighted individual spectra. The error vectors are created by co-adding the individual inverse variance vectors. The final errors on the line measurements from the stacked spectra are the actual line measurement errors.
The line measurements from the stacked spectra are compared to the error-weighted means of the local sample. The errors on the $z=0$ points represent the error-weighted error of the mean. 

The Balmer decrements measured from the spectra stacked as a function of magnitude are shown in 
Figure \ref{fig:stack_j}. Out of the total sample of 341 objects at $0.78<z<0.83$, the J-band stacks include 174 objects. AGN and AGN+SF objects are also exluded. The number of objects in each bin, the line fluxes, line S/N and attenuations in each bin are listed in Table \ref{tab:data1}. The points indicate a shallow decrease of the attenuation with increasing magnitude however the correlation does not have a high statistical significance -- even without the brightest bin the Pearson coefficient is $-0.31$. The most luminous bin exhibits attenuation significantly below the local value. The large $\chi^{2}=91$ of the difference between the z=0.8 and z=0.0 samples is mostly driven by this bin. It is possible that the most luminous bin suffers from AGN contribution from objects which cannot be individually identified as AGN because they have too low line S/N to be placed on the diagnostic diagram. The individual stacked spectra are shown in Figure \ref{fig:stacked_spectra}. The Balmer lines of the stacked spectrum in the first bin are slightly broader than those of the fainter bins suggesting that AGN contamination may indeed be present.

\begin{figure}
\figurenum{15}
\label{fig:stack_mass}
\plotone{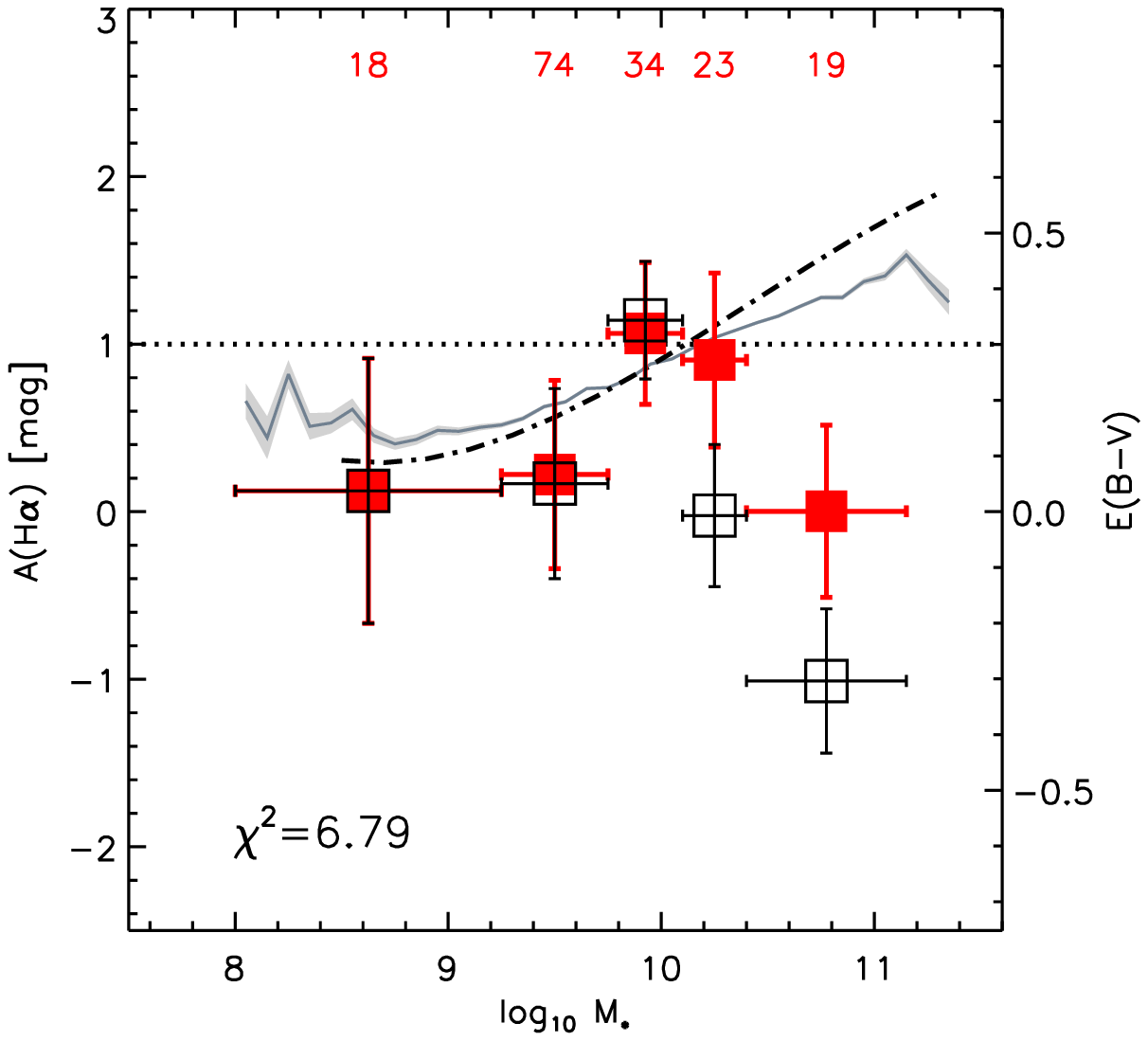}
\caption{Mean attenuation based on the Balmer decrements measured from stacked spectra, binned by the logarithm of the stellar mass. The symbols are the same as in Figure \ref{fig:stack_j}. The local relation between attenuation and stellar mass \citep{garn10} is also plotted (dot-dash line). The weighted mean attenuation for the local \hg/\hb\ SDSS sample as a function of stellar mass is shown (solid dark gray line) with errors indicating the weighted error on the mean (light gray shading). $\chi^2=6.79$ is a measure of the differences between the two samples. Figure \ref{fig:stacked_spectra_mass} shows the binned spectra and Table \ref{tab:data2} lists the line fluxes and attenuation values for each bin.}
\end{figure}

\begin{figure*}
\figurenum{16}
\label{fig:stacked_spectra_mass}
\includegraphics[angle=0,width=6.5in]{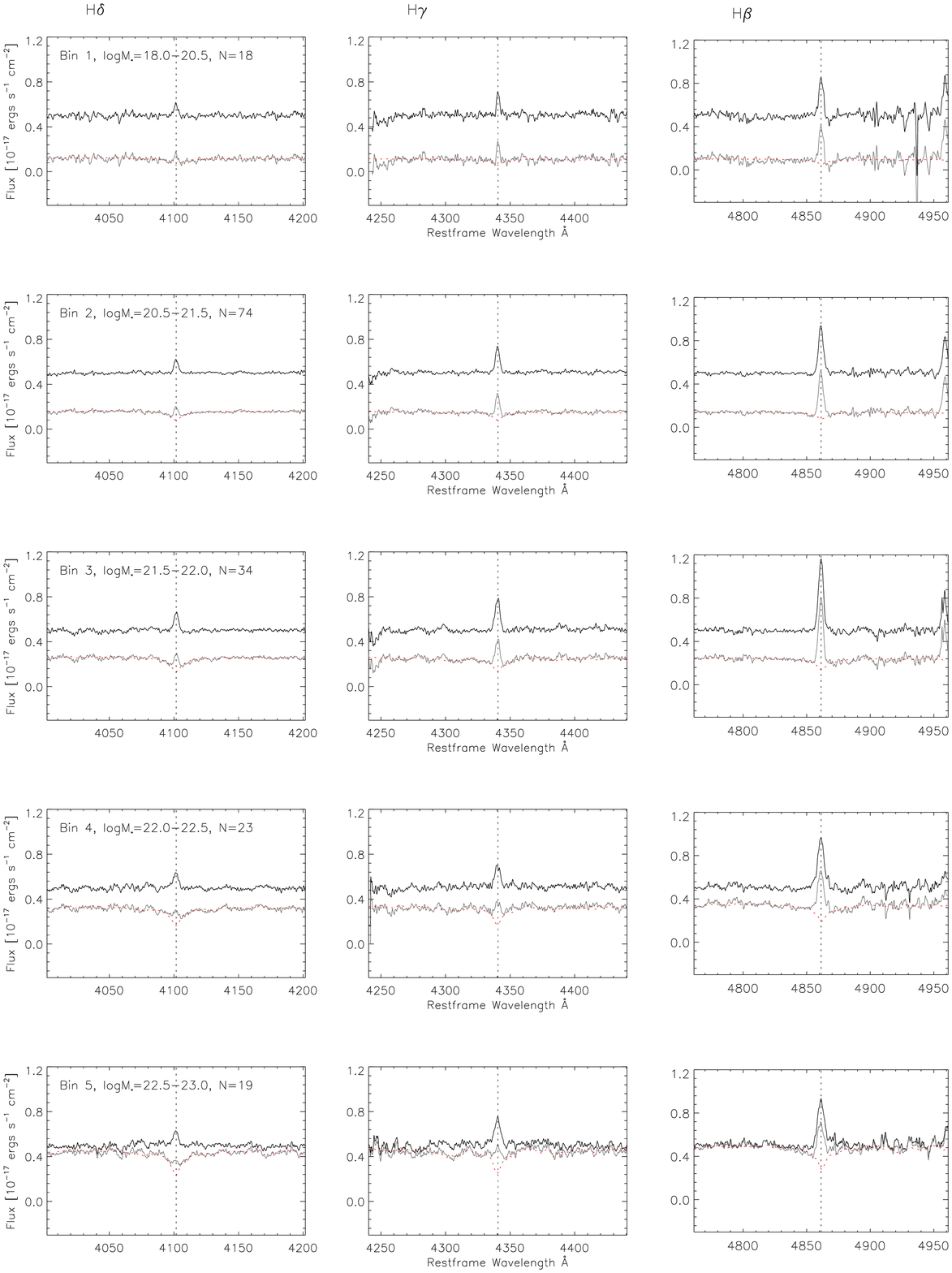}
\caption{Same as Figure \ref{fig:stacked_spectra} but for the logarithmic mass bins.}
\end{figure*}

A number of studies in the local universe have shown that attenuation increases in star-forming galaxies with stellar mass, such that more massive galaxies are dustier. Figure \ref{fig:stack_mass} shows the stacked Balmer decrement measurements as a function of the stellar mass. A total of 168 individual spectra are combined in the stacks. The remaining objects are excluded due to lack of mass measurements and prominent sky lines in the spectra. The attenuation shows a mild increase towards higher masses, more prominent in the \hg/\hb\ sample. Again, AGN contamination may be present in the most massive bin where $M_{\ast}>10.4$, as expected from Figure \ref{agn}. The $\chi^{2}=7$ is mainly dominated by this bin. The remaining mass bins closely follow the local relation.
The stacked spectra are shown in Figure \ref{fig:stacked_spectra_mass} and the line measurements for each bin are presented in Table \ref{tab:data2}.

\begin{figure}
\figurenum{17}
\label{fig:stacknb}
\plotone{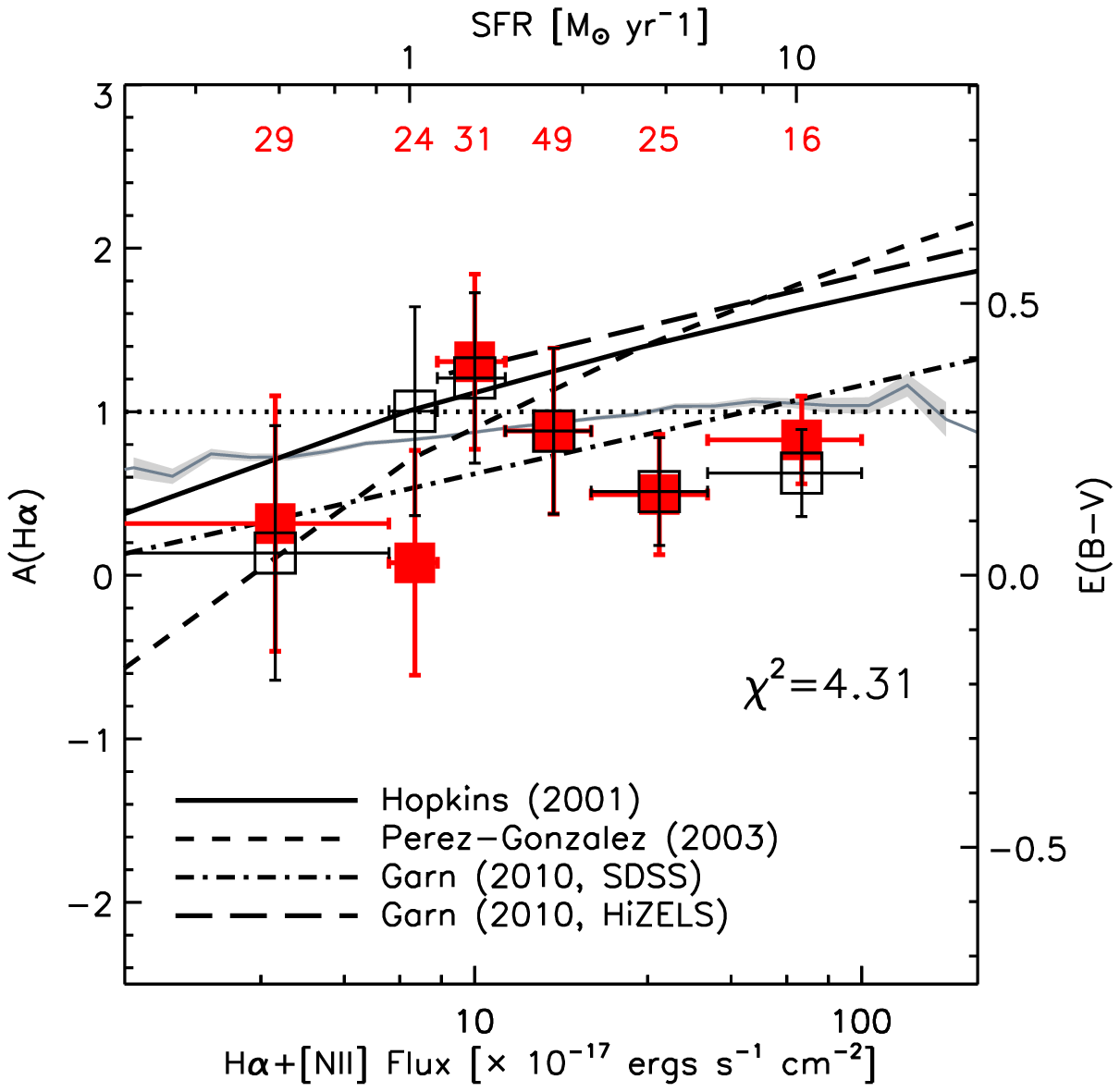}
\caption{Mean attenuation based on the Balmer decrements measured from stacked spectra, binned by their $H\alpha+$[NII]$\lambda$6584 flux. The symbols are the same as in Figure \ref{fig:stack_j}.  The weighted mean attenuation for the local \hg/\hb\ SDSS sample as a function of stellar mass is shown (solid dark gray line) with errors indicating the weighted error on the mean (light gray shading). $\chi^2=4.31$ is a measure of the differences between the two samples. The local relations of \citet{pg03}, \citet{hopkins01}, and \citet{garn09} as well as the $z=0$ relation of \citet{garn10} are also included as indicated by the legend. Figure \ref{fig:stacked_spectra_nb} shows the binned spectra and Table \ref{tab:data3} lists the line fluxes and attenuation values for each bin.}
\end{figure}

\begin{figure*}
\figurenum{18}
\label{fig:stacked_spectra_nb}
\includegraphics[angle=0,width=6.5in]{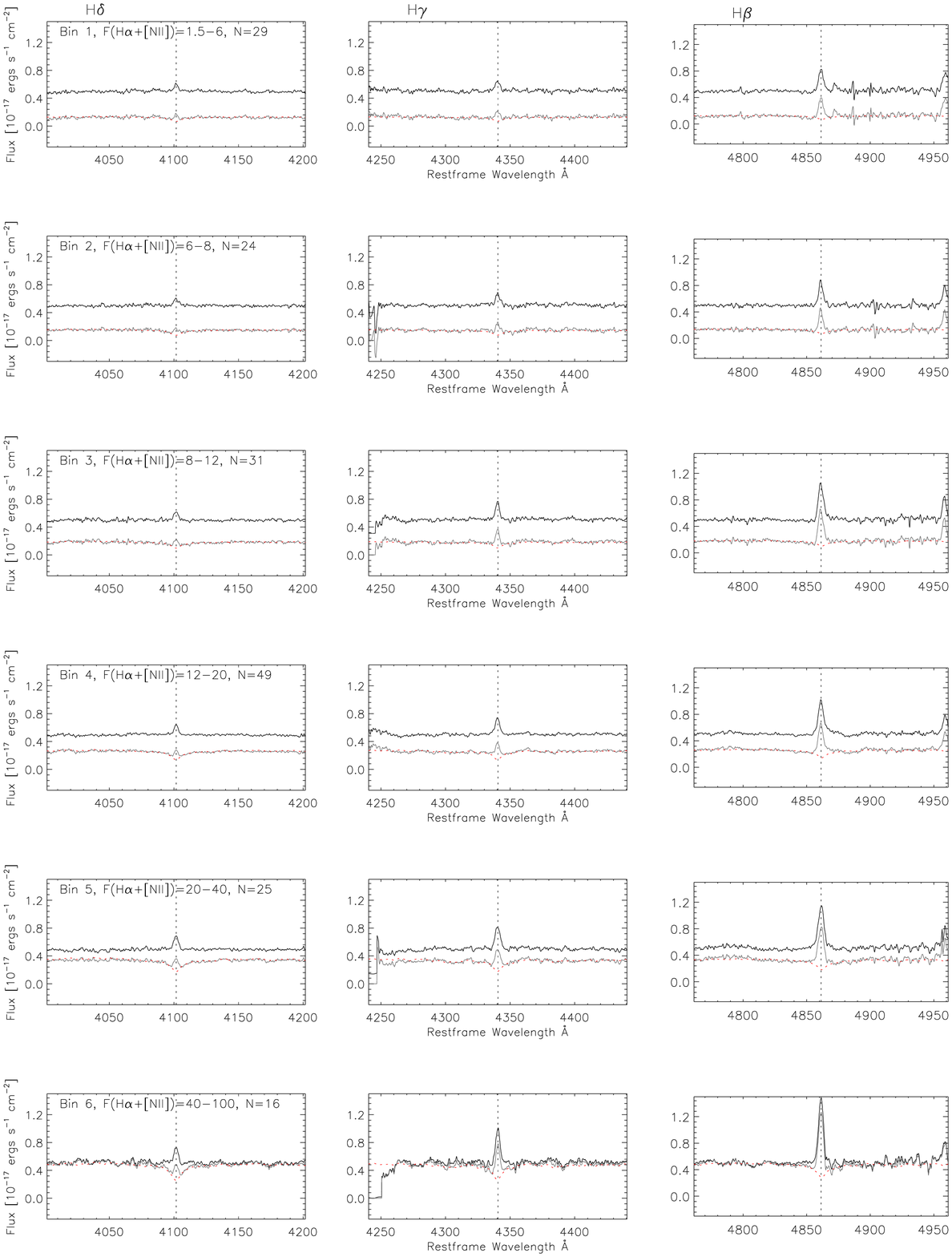}
\caption{Same as Figure \ref{fig:stacked_spectra} but for the binning in NB1190 flux in units of $10^{-17}$ erg s$^{-1}$ cm$^{-2}$.}
\end{figure*}

Finally, local studies indicate that dust content and attenuation increase with the observed SFR \citep[e.g.,][]{hopkins01, sullivan01,garn10} so the same procedure is carried out as a function of $H\alpha+$[NII]$\lambda$6584 flux. The results are presented in Figure \ref{fig:stacknb}. 174 emitters are used in these stacks. The attenuation in the stacked spectra does slowly increase with $H\alpha+$[NII]$\lambda$6584 flux, in good agreement with the running weighted mean of the SDSS sample. The good agreement with the local sample is measured by the low $\chi^{2}=4$ of the difference between the two samples.
The stacked spectra and line measurements are shown in Figure \ref{fig:stacked_spectra_nb} and Table \ref{tab:data3}, respectively. 

Figures \ref{fig:stack_j}, \ref{fig:stack_mass} and \ref{fig:stacknb} also show a comparison between the stacked spectra which include only star-forming galaxies and stacked spectra which also include the AGN and AGN+SF objects identified in Section 5. The line ratios of the stacked spectra which include AGN and composite galaxies are overall consistent in most bins with the stacks which do not. However, greater than $1\sigma$ differences appear in the most massive and most luminous bins, such that the stacks which include AGN have lower attenuation. In the remaining bins AGN which were not individually identified should have a minor effect.

 \section{Discussion}

In the previous sections we presented Balmer decrement measurements for a sample of \ha-selected star-forming galaxies at $z=0.8$. Both the individual and stacked spectra show that the average attenuation of the sample is $A(H\alpha)\sim1$ magnitude. The attenuation as a function of both stellar mass and SFR is similar to those at $z=0$ in comparison to the SDSS sample. In this section we compare these results to those from other local and intermediate-redshift samples and discuss the possible limitations of our analysis. 

\subsection{Comparison with Other Extinction Relations}

Previous studies have found a correlation between mass and attenuation in star-forming galaxies. \citet{brinchmann04} notice a clear increase in dust attenuation as a function of mass as well as a wider range of attenuations for massive galaxies. For the SDSS sample they find that the mass-weighted attenuation $A(H\alpha)$ is 1.3 with a spread of $\sim0.7$ magnitudes. Based also on SDSS and using a principal component analysis, \citet{garn10} claim that mass is the most fundamental property which correlates with the attenuation, while SFR and metallicity are secondary, mainly brought about by the dependence of these parameters on mass.  \citet{garn09} compare SFR indicators for a sample of $z=0.84$ \ha-selected emitters similar to ours, {estimating the attenuation from the ratio of the H$\alpha$ and 24$\mu m$ fluxes}. They find that the mass-attenuation relation at that redshift is similar to the local one, with a possible  plateau at $M_{\ast}>1.6\times10^{10}M_{\odot}$. With the current sample we also find that $z=0.8$ galaxies show a mass-attenuation relation similar to the local one. In Figure \ref{fig:stack_mass} we compare our results with the derived relation of \citet{garn10}, based on the SDSS sample:
\begin{equation}
A(H\alpha) = 0.91 + 0.77X + 0.11X^{2} - 0.09X^{3},
\end{equation}
where $X=\log_{10}(M_{\ast}/10^{10}M_{\odot})$. We find agreement within the uncertainties, indicating little or no evolution  in the amount of attenuation at a given stellar mass over the last 6.8 Gyrs. A comparison with the \citet{garn09} relation for their sample of $z=0.84$ emitters (their Figure 6b) also shows consistency, despite the different methods for determining the attenuation. 

The relation between SFR (or \ha\ luminosity) and attenuation is frequently studied in light of efforts to derive attenuation-corrected SFR with minimum auxiliary information. A number of previous studies \citep[e.g., ][]{hopkins01, sullivan01, dopita02, pg03, schmitt06, caputi08} have found such a correlation locally, with the implication that galaxies with higher SFRs are also dustier. Most such works have found a generally consistent slope and normalization. Here, we compare our results with the following two local relations:
\begin{eqnarray}
\log SFR &=& (0.597\pm0.021)\frac{H\alpha}{H\beta}+(-2.191\pm0.086), \nonumber \\  
& &\mathrm{(P{\acute{e}}rez-Gonz{\acute{a}}lez\ et\ al.\ 2003,\ Eq.4)}
\end{eqnarray}
\begin{eqnarray}
SFR_{o} &=& SFR_{i}-2.614\times\log_{10}\left(\frac{0.797\times SFR_{i}+3.834}{2.88}\right), \nonumber \\ 
& &\mathrm{(Hopkins\ et\ al.\ 2001,\ Eq.5)} 
\end{eqnarray}
The \citet{pg03} study is based on a sample of nearby objective-prism-selected star-forming objects. The SFR is derived from the observed \ha\ luminosity, using the \citet{sullivan01} conversion. The \citet{hopkins01} relation is based on the \citet{wang96} sample of $z\sim0$ FIR-detected galaxies and shown to be a good fit to a large ($\sim500$ objects) compilation of local star-forming galaxies from \citet{cram98}. The relations between attenuation and star-fromation have been extended up to $z=0.84$ by \citet{garn09} {(albeit using a different attenuation indicator)}, in agreement with local ones:
\begin{equation}
A_{H\alpha} = -21.963+0.562\log_{10}\frac{L_{H\alpha, obs}}{erg\ s^{-1}}
\end{equation}
Our $z=0.8$ sample is in agreement with both the local and the \citet{garn09} relation as shown in Figure \ref{fig:stacknb}.

Both comparisons show that the dust content of star-forming galaxies as a function of stellar mass and SFR have not significantly changed during the second half of the lifetime of the Universe. This finding appears to be at odds with studies which have measured increased attenuation in galaxies at $0.5<z<1.0$ \citep[e.g,][]{flores04, liang04, rodrigues08,villar08,savaglio05}. However, the majority of such studies have targeted LIRGs and ULIRGs. While a detailed comparison between the attenuation of LIRG and non-LIRG galaxies within our sample is beyond the scope of this paper, we can briefly examine the issue here. Deep 24 $\mu$m MIPS/Spitzer observations of a portion of our SXDS field have been acquired by the Spitzer UKIDSS Ultra Deep Survey \citep[SpUDS,][]{spuds}, reaching $3\sigma$ flux level of 25 mJy. This limit corresponds to a total IR luminosity (TIRL) $L_{IR} = 2\times10^{10}\ L_{\odot}$ for an object at $z=0.8$ using the \citet{rieke09} relation between observed 24 $\mu$m flux and TIRL. A total of 50 \ha\ emitters with spectroscopic redshifts fall within the SpUDS footprint. We match 24 $\mu$m counterparts to 30 of these emitters (using a 3\arcsec\ matching radius) and find that 40\% of the sample in the field (20 objects) have LIRG-like TIR luminosities ($L_{IR} >10^{11} \ L_{\odot}$).  Only five of the LIRGs have measured Balmer decrements so, instead of using the decrements, we derive A(\ha) by comparing the SFR derived from the \ha\ flux and from the 24 $\mu$m flux in a manner similar to  \citet{garn09}. We find that the median attenuation of the non-LIRG and LIRG galaxies is A(\ha)$=1.0\pm0.8$ and $1.8\pm0.7$ magnitudes, respectively, indicating a much higher attenuation for the LIRGs, consistent with prior studies. Within our sample, the LIRGs have brighter J-band magnitudes and dominate the sample for $J<21.5$. The LIRG galaxies also have higher \ha$+$[NII]$\lambda$6584 fluxes: the median \ha$+$[NII]$\lambda$6584 flux for the LIRG sample is 2.7 times higher than the median of the 30 24$\mu$m non-detections and non-LIRG sources: $20.5\times10^{-17}$ vs.  $7.6\times10^{-17} ergs\ s^{-1}\ cm^{-2}$. The bright J-band magnitudes and elevated median \ha\ flux indicate that the LIRGs in our sample occupy the parameter space where we have already found elevated attenuation (Figures \ref{fig:stack_j} and \ref{fig:stack_mass}), and have a smaller contribution to our sample at low-luminosity and low-SFR.

However, elevated attenuations are also found in studies of galaxies with lower SFRs. For example, in a study of 69 $0.4<z<1.0$ galaxies in the spectroscopic Gemini Deep Deep Survey \citep{abraham04}, \citet{savaglio05} find \hg/\hb$=0.29\pm0.03$ and \hd/\hb$=0.18\pm0.02$ in a stacked spectrum of {\it all} objects, lower than most individual detections in our sample (corresponding to A(\ha) of $3.34\pm0.2$ and $1.68\pm0.15$, respectively). \citet{villar08} also find higher attenuation in their narrow-band \ha-selected $z=0.84$ sample -- $A(H\alpha)=1.48$.  While our results are consistent with this value within the errors, we do not find a shift in the average attenuation of galaxies with physical properties similar to the local sample. {The difference may be due to the different attenuation measurement method -- instead of Balmer decrements \citet{villar08} use determine the mean attenuation for their sample via the $\mathrm{F}_{dust}/\mathrm{F}_{FUV}$ ratio \citep{buat05}. If the FUV continuum, the Balmer lines and the dust emission sample different physical scales, the attenuations may differ. Nevertheless, a comparison  is valuable, because their sample selection is very similar to ours. Similar to their work, we derive a single attenuation for our full sample, to determine if  their average is weighted towards more massive and luminous (and more attenuated) sources.  }However, even if, instead of splitting our sample in bins by SFR, we combine them in a single stacked spectrum, we find {\it lower}, not higher values of the attenuation: A(\ha)=$0.25\pm0.75$ from the ratio of \hg\ to \hb\, and A(\ha)$=0.55\pm0.65$ from the ratio of \hd\ to \hb. These values are $\sim0.5$ magnitude lower than the median attenuation from the individual detections. They are also $\gtrsim1$ magnitude lower than the values of \citet{villar08} and inconsistent within $1\sigma$ with the measurements by \citet{savaglio05}. The stacked spectrum is dominated by low-luminosity galaxies which in both SDSS and in New\ha~ have attenuation of $<1$ magnitude. Even if we weigh equally all galaxies in the manner of \citet{savaglio05} who normalize all spectra to one before stacking them, we still measure attenuation of A(\ha)$\sim1.0$ magnitude, inconsistent with their results. In conclusion, we do not find evidence for systematically elevated attenuation relative to the local sample.

\subsection{Limitations of the Balmer Decrement Analysis}

Problems may exist with our approach to determining the attenuation using the high order Balmer decrements. In this section we discuss the limitations of using \hb, \hg\ and \hd\ to constrain the attenuation.

The high-order Balmer lines are faint. \hg\ is 5.85 times fainter than \ha\ (assuming Case B), and \hd\ another factor of two fainter. We have shown that this can be overcome by deep spectroscopic observations, relatively high spectral resolution and stacking of sources. Assuming Case B for the \hg/\hb\ ratio, we can calculate that in order to achieve an error in A(\ha) of 0.5 magnitudes both lines must be detected at a S/N$>20$. To achieve the same precision with \hd/\hb\ we need $S/N>14$. In comparison, the median error of the individual detections of our sample is 0.7 magnitudes. There are only seven objects with \hg\ detected with S/N$>20$ and six objects with \hd\ detected with S/N$>14$, all of them identified as AGN or composite objects in Section 5. The mean Balmer ratios for these two samples are $0.42\pm0.04$ and $0.22\pm0.03$ for \hg/\hb\ and \hd/\hb, respectively, consistent with our findings for the full samples. Furthermore, by stacking individual spectra we have been able to measure the lines with S/N$\gtrsim15$, again achieving consistent results. 

A complementary line of work would be to construct the \ha/\hb\ line ratio using the \ha\ fluxes from our NEWFIRM narrow-band NB118 photometry. Assuming that the continuum noise is the same at all wavelengths, the S/N in the \ha\ line will be $\sim14$ times higher than that of \hb\ and $\sim30$ times higher than that of \hg\ (Case B). Such an analysis requires a careful examination of the possible sources of systematic error when combining spectroscopic and photometric data such as differences between the apertures within which the \hb\ and \ha\ line fluxes are measured. Furthermore, the NB118 bandpass is wide enough to include flux from the [N II] $\lambda\lambda6548,6583$ emission lines for the \ha\ emitters and we need to correct the \ha\ fluxes for the contamination in order to construct the \ha/\hb\ decrement. The \ha/[NII]$\lambda$6584 ratio is a metallicity indicator and thus varies as a function of equivalent width \citep{villar08}, B-band luminosity \citep{kennicutt08,lee09} and mass, but the scatter in these relations is significant. Near-IR spectroscopy is required to directly measure the H$\alpha$/[NII]$\lambda$6584 ratio.

In addition to the measurement sources of error, there are physical weaknesses in using Balmer decrements to determine attenuation. One regime where the optical attenuation measurements break down is in extremely dusty galaxies. Even though many such objects may exhibit an emission line spectrum indicating moderate attenuation, the nebular reddening may bear little resemblance to the actual dust content. In such cases the emission comes from the least-obscured regions of the galaxy and is not representative of the galaxy as a whole. This issue is manifested in observed discrepancy between the Balmer decrement attenuation and IR-to-\ha\ attenuation. In such galaxies, the Balmer decrements underestimate the actual attenuation. The mid-IR and far-IR emission is much more sensitive in unmasking obscured star formation and can be used to address this issue in future work.

\section{Summary and Future Work}

We have examined the dust attenuation in normal star forming galaxies at $z=0.8$ using the \hg/\hb\ and \hd/\hb\ line ratios. We have carried out deep spectroscopic observations for a sample of narrow-band \ha\ selected galaxies, which allows us to detect \hb, \hg\ and \hd\ with S/N$>5$. We apply several different techniques to isolate AGN and composite objects from the sample: the MEx diagnostic, the UV variability and the spectral type. In total, we identify 17 AGN and seven composite objects which indicates a strict lower limit on the AGN fraction of 5\%, in line with other studies.

We find good agreement between the distribution of attenuations in local galaxies and those from the New\ha\ sample. The mean attenuation of galaxies with individual line detections is $A(H\alpha)=0.8\pm1.0$ and $0.7\pm1.0$ for the \hg/\hb\ and \hd/\hb\ samples respectively. The mean (median) of the joint sample is A(\ha)$=0.8(0.9)\pm1.0$.

We stack the spectra in bins of J-band magnitude, stellar mass and \ha$+$[NII]$\lambda$6584 flux in order to increase the S/N in the spectra and probe a more representative sample of the overall star-forming population than possible with individual detections. The stacked spectra also indicate consistency between the attenuation at $z=0.8$ and $z=0$ for galaxies with the same mass and SFR. 

The analysis presented here is  solely focused on the Balmer decrement as an attenuation indicator. As we have pointed out, this method is not applicable for very dusty objects where the nebular emission originates only in the least dusty, optically thin regions of the galaxy. In order to explore this regime, we plan to use the MIPS 24$\mu$m observations from the SpUDS survey which overlaps with part of our SXDS field. Further work will also include comparison of the Balmer decrement results with measurements of attenuation from the UV slope and SED models.

\acknowledgements
We thank Jarle Brinchmann for providing assistance with the MPA/JHU catalogs and Brent Groves for sharing his preliminary results on the SDSS Balmer decrements. We thank Stephanie Juneau for providing us with results regarding the MEx diagnostic. We also thank Christy Tremonti
for providing the spectral fitting code. The NewH$\alpha$ Survey has been primarily funded by
Hubble and Carnegie Fellowships to J.C.L., and we are grateful for the support from these programs. C.L. was supported by NASA grant NNX08AW14H through their Graduate Student Research Program.

Based on observations made with the NASA Galaxy Evolution Explorer.
{\it GALEX} is operated for NASA by the California Institute of Technology under NASA contract NAS5-98034.

{\it Facilities:} Magellan-2 (IMACS), Mayall (NEWFIRM), Blanco (NEWFIRM), Subaru (Suprime-Cam), GALEX

\appendix{}

\section{Appendix: Balmer Decrements from Optical Spectroscopy}

The Balmer emission lines arise from the recombination and subsequent cascade of electrons to the $n=2$ level of the hydrogen atoms in the interstellar medium. The preceding photoionization is caused by the radiation field of hot, young, massive stars. Collisional excitation does occur as well \citep[e.g., ][]{ferland09} but photoionization and recombination are the dominant processes.

The simple atomic structure of hydrogen has made it possible to determine the expected rates of transition and the emission line ratios as a function of the temperature and electron density of the interstellar medium. Two general cases are considered when discussing the Balmer line emissons. In the Case A recombination, the nebula is optically thin to the photons of the Lyman series (i.e., photons emitted from transitions to the $n=1$ level of hydrogen). Alternatively, in the Case B recombination model, the nebula is optically thick to photons of the Lyman series but optically thin to Ly$\alpha$ and photons of the higher hydrogen series. Case B is typically assumed for determining the intrinsic line ratio. Here we assume Case B recombination and electron density and temperature typical for HII regions: $10^4$ cm$^{-2}$ and $10,000$ K. The expected intrinsic Balmer line ratios for these condition are \citep{osterbrock89}:
\begin{eqnarray}
(H\alpha/H\beta)_{int}&=&2.86\\
(H\gamma/H\beta)_{int}&=&0.469\\
(H\delta/H\beta)_{int} &=& 0.260
\end{eqnarray}

The ratios are only weakly sensitive to electron density and more sensitive to temperature. For example the $H\alpha/H\beta$ ratio varies by only 0.05 (2.86 to 2.81) for electron densities $n_{c}=10^{2}$ to $10^{6}\ cm^{-3}$ but by 0.29 for temperatures between 5000K and 20,000K \citep{dopita}. However these variations are still small relative to the effects of dust. 

Our optical spectra only span the 3500 to 5300\AA~ region and do not contain the \ha~ emission line at the redshift of the sample. We do have \ha\ observations from our NB1190 photometry but the direct comparison between the spectroscopic and photometric line fluxes is impeded by the need of aperture corrections which we defer to future work. In this paper we focus on the $H\gamma/H\beta$ and $H\delta/H\beta$ decrements. The effect of dust is to diminish the observed emission of the short-wavelengths more than the long wavelengths. The attenuated observed line ratios will then be   smaller than the intrinsic ones due to the effect of dust. The color excess due to this dust attenuation can be expressed as:

\begin{eqnarray}
A(H\beta) - A(H\gamma) &=&\\
 = E(H\beta - H\gamma) &=& -2.5\times\log_{10} \left[ \frac{(H\gamma/H\beta)_{int}}{(H\gamma/H\beta)_{obs}} \right],
\end{eqnarray}

where $(H\gamma/H\beta)_{obs}$ denotes the observed line ratio. This color excess can be related to the broadband color excess $E(B-V)$ via an attenuation curve:

\begin{equation}\label{eqa6}
A(\lambda) = \kappa(\lambda)E(B-V),
\end{equation}

where $\kappa(\lambda)$ is the value of the attenuation curve at the wavelength $\lambda$. Using the attenuation curve we can express the attenuation at $H\gamma$ and $H\beta$ in the equation above:

\begin{equation}
E(H\beta - H\gamma) = E(B-V)[\kappa(H\beta) - \kappa(H\gamma)]
\end{equation}

The functional form of the attenuation curve is a matter of choice. We follow the common practice and adopt the \citet{calzetti97} attenuation curve in the form:

\begin{eqnarray}
\kappa'(\lambda) &=& 2.659(-1.857+1.040x)+R_{v}', \\
& &\mathrm{for}\ 0.63\mu m\leq \lambda \leq 2.20\mu m;\\
&=&2.659(-2.156+1.509x-0.198x^{2}+0.011x^{3}) +R_{v}',\\
& &\mathrm{for}\ 0.12\mu m\leq \lambda \leq 0.63\mu m,\\
\end{eqnarray}
where $x=1/\lambda$ is the wave number and $R_{v}'=A(V)/E(B-V)=4.05$ is the ratio of the total to selective attenuation. Using this prescriptions, we find $\kappa(H\alpha)=3.33$, $\kappa(H\beta)=4.60$ and $\kappa(H\gamma)=5.12$ and $\kappa(H\delta)=5.39$. The expression for $E(H\beta - H\gamma)$ in Equation 7 can be then substituted in Equation 5, reaching a solution for the broadband color excess $E(B-V)$:

\begin{eqnarray}
E(B-V) &=& \frac{E(H\beta - H\gamma)}{\kappa(H\beta)-\kappa(H\gamma)} \\
&= &\frac{-2.5}{\kappa(H\beta)-\kappa(H\gamma)}\times \log_{10} \left[ \frac{0.469}{(H\gamma/H\beta)_{obs}} \right]
\end{eqnarray}
Similarly, the broadband color excess can be expressed as a function of the $H\delta/H\beta$ Balmer decrement is:

\begin{eqnarray}
E(B-V) &=& \frac{E(H\beta - H\delta)}{\kappa(H\beta)-\kappa(H\delta)} \\
&= &\frac{-2.5}{\kappa(H\beta)-\kappa(H\delta)}\times \log_{10} \left[ \frac{0.260}{(H\delta/H\beta)_{obs}} \right]
\end{eqnarray}
From here we can express the attenuation at an arbitrary wavelength via Eq. \ref{eqa6}. Most commonly the attenuation is determined at $\lambda_{V}=5500$\AA, where $\kappa(V) = 4.05$. For the purposes of our work, we are also determine $A(H\alpha)$, the attenuation at \ha\ ($\lambda=6563$\AA), where $\kappa(H\alpha) = 3.33$.

\clearpage

\begin{deluxetable}{lllcccccc}
\tablecolumns{7}
\tablewidth{0pc}
\tabletypesize{\footnotesize}
\tablecaption{AGN and Composite Candidates \label{tab:agn}}
\tablehead{
\colhead{New\ha} & \colhead{RA} & \colhead{Dec} & \colhead{FUV}    & \colhead{Spectral} & \colhead{MEx\tablenotemark{a}} &  \colhead{Comment}
\\
\colhead{Field} & \colhead{} &  \colhead{} &  \colhead{Variability} &  \colhead{Type} & \colhead{} &   \colhead{}
}
\startdata

SXDSN & 02:18:22.5  & -04:30:36.1 & + & + & -  & - \\ 

SXDSS  & 02:17:42.6  & -05:03:51.4 &- & + & -  & BL\tablenotemark{b} \\ 
SXDSW & 02:17:01.3  & -04:54:52.6 & - & + & -  & - \\ 
SXDSW & 02:16:14.7  & -04:46:05.3 & - & + & -  & BL AGN \\ 
SXDSN & 02:17:43.0  & -04:36:25.0 & - & + & -  & BL AGN \\ 

SXDSS & 02:17:21.9  & -05:03:49.2 & - & - & Sy2 (0.412)/L(0.324)   & -\\ 
SXDSS & 02:16:52.5  & -05:22:30.5 & - & - & L (0.487) & - \\ 
SXDSS & 02:18:20.1  & -05:11:53.3  & - & -& L(0.5)/Sy2 (0.5)  & - \\ 
SXDSS & 02:16:50.9  & -05:21:15.1  & - & -& L (0.561)  & - \\ 
SXDSW & 02:17:09.9  & -04:55:56.7 & - & - & Sy2 (0.667) & - \\ 
SXDSW & 02:16:31.7  & -04:57:25.5 & - & - & Sy2 (0.857) & - \\ 
SXDSW & 02:15:51.8  & -05:09:47.7 & - & - & L (0.341)/Sy2 (0.317) & - \\ 
SXDSW & 02:17:16.3  & -04:59:57.4 & - & - & L (0.519) & - \\ 
SXDSN & 02:18:16.6  & -04:26:34.4  & - & -& Comp (0.5 )/Sy2 (0.5)  & - \\ 

SXDSS & 02:17:37.1  & -05:06:22.2 & - & + & Comp (0.610)  & - \\ 
SXDSS & 02:17:21.8  & -05:17:24.4 & - & -& Comp (0.504)  & - \\ 
SXDSS & 02:16:45.2  & -05:18:03.5 & - & + & Comp (0.570) & - \\ 
SXDSS & -05:16:48.4 & -05:18:42.5 & - & -& Comp (0.560) & - \\ 
SXDSW & 02:16:36.8  & -05:05:54.0 & - & - & Comp (0.802)  & - \\ 
SXDSW & 02:15:48.4  & -04:53:55.9 & - & - & Comp (0.723)  & - \\ 
SXDSW & 02:16:57.9 & -05:05:39.5 & - & -& Comp (0.467) & - \\ 

SA22 & 22:17:30.8 & +00:12:18.9  & - & - & + & - \\ 
SA22 & 22:17:32.5 & +00:21:33.4  & - & - & + & - \\ 
SA22 & 22:17:12.0  & +00:12:44.1 & - & - & + & - \\  


\enddata
\tablenotetext{a}{Classification according to \citet[][see Figure \ref{agn}]{juneau}. The designations indicate LINER (L), Seyfert Type 2 (Sy2) and Composite SF+AGN (Comp). The fractional probability that the object belongs to a given class is given in brackets. Objects are assigned the class which has the largest fractional probability. The probabilities of the four classes (L, Sy2, Comp and SF) add up to 1.}
\tablenotetext{b}{Broad line.}
\end{deluxetable}

\clearpage

\begin{deluxetable}{lcccclllllll}
\tablecolumns{12}
\tablewidth{0pc}
\tabletypesize{\footnotesize}
\tablecaption{Stacked Spectra as a Function of J-band Magnitude \tablenotemark{a}\label{tab:data1}}
\tablehead{
\colhead{Bin}  & \colhead{J[mag]} & \colhead{N\tablenotemark{b}}  & \colhead{Mean S/N\tablenotemark{c}}    & \colhead{\hb\tablenotemark{d}} & \colhead{$\Delta$\hb}& \colhead{\hg} & \colhead{$\Delta$\hg}& \colhead{\hd}& \colhead{$\Delta$\hd} & \colhead{A(\hg/\hb)} & \colhead{A(\hg/\hb)}}
\startdata
1  &  19.45 - 20.75  &  16  & 3.7  &   3.12  &   0.15  &    1.65  &   0.12   &   0.95  &  0.09  & -0.82 & -1.1\\
2  &  20.75 - 21.25  &  24  & 2.8  &   3.31  &   0.15  &    1.29  &   0.12   &   0.71  &   0.09  & 1.25 & 1.29\\
3  &  21.25 - 21.75  &  28  & 2.1  &   3.45  &  0.16   &    1.57  &   0.10    &  0.78   & 0.08   & 0.20 & 0.93 \\
4  &  21.75 - 22.25  &  33  & 1.6  &   2.33  &   0.14  &    1.03  &   0.10   &  0.52   & 0.07   & 0.43  & 0.99\\
5  &  22.25 - 22.80  &  35  & 1.2  &   2.19  &   0.13  &    0.94  &   0.10   &  0.44   & 0.08   & 0.60  & 1.82\\
6  &  22.80 - 25.00  &  38  & 0.9  &   2.00  &   0.12  &    0.84  &   0.10   &  0.48   & 0.07   &0.75  & 0.53 \\
\enddata
\tablenotetext{a} {\ The data presented here are for the stacked spectra which exclude individually-identified AGN and composite sources.}
\tablenotetext{b} {\ Number of individual spectra used to create the stacked spectrum.}
\tablenotetext{c} {\ Mean continuum signal-to-noise in the stacked spectrum.}
\tablenotetext{d} {\ Line fluxes are in units of $10^{-17}$ ergs s$^{-1}$ cm$^{-2}$. They are not corrected for attenuation or aperture losses.}
\end{deluxetable}

\begin{deluxetable}{lcccclllllcc}
\tablecolumns{12}
\tablewidth{0pc}
\tabletypesize{\footnotesize}
\tablecaption{Stacked Spectra as a Function of Stellar Mass\tablenotemark{a} \label{tab:data2}}
\tablehead{
\colhead{Bin}  & \colhead{$M_{\ast}$} & \colhead{N\tablenotemark{b}}  & \colhead{Mean S/N\tablenotemark{c}}    & \colhead{\hb\tablenotemark{d}} & \colhead{$\Delta$\hb}& \colhead{\hg} & \colhead{$\Delta$\hg}& \colhead{\hd}& \colhead{$\Delta$\hd} & \colhead{A(\hg/\hb)} & \colhead{A(\hg/\hb)}}
\startdata
1 & 8.00 - 9.25 		& 18 & 0.8 & 1.48 & 0.14 & 0.69 & 0.10 & 0.37 & 0.08 & 0.10 & 0.22 \\
2 & 9.25 - 9.75 		& 74 & 1.2 & 2.01 & 0.13 & 0.94 & 0.10 & 0.48 & 0.07 & 0.03 &  0.62\\
3 & 9.75 - 10.10 	& 34 & 1.9 & 3.30 & 0.16 & 1.30 & 0.10 & 0.70 & 0.08 & 1.19 & 1.41\\
4 & 10.10 - 10.40 	& 23 & 2.4 & 2.86 & 0.16 & 1.10 & 0.11 & 0.68 & 0.09 & 1.39 & 0.63\\
5 & 10.40 - 11.15 	& 19 & 3.3 & 2.82 & 0.16 & 1.39 & 0.13 & 0.68 & 0.10 & -0.35 & 0.53\\
\enddata
\tablenotetext{a} {\ The data presented here are for the stacked spectra which exclude individually-identified AGN and composite sources.}
\tablenotetext{b} {\ Number of individual spectra used to create the stacked spectrum.}
\tablenotetext{c} {\ Mean continuum signal-to-noise in the stacked spectrum.}
\tablenotetext{d} {\ Line fluxes are in units of $10^{-17}$ ergs s$^{-1}$ cm$^{-2}$. They are not corrected for attenuation or aperture losses.}
\end{deluxetable}

 \begin{deluxetable}{lcccclllllll}
\tablecolumns{12}
\tablewidth{0pc}
\tabletypesize{\footnotesize}
\tablecaption{Stacked Spectra as a Function of $H\alpha+$[NII]$\lambda$6584 Flux \tablenotemark{a} \label{tab:data3}}
\tablehead{
\colhead{Bin}  & \colhead{\ha$+$[NII]\tablenotemark{b}} & \colhead{N\tablenotemark{c}}  & \colhead{Mean S/N\tablenotemark{d}}    & \colhead{\hb\tablenotemark{b}} & \colhead{$\Delta$\hb}& \colhead{\hg} & \colhead{$\Delta$\hg}& \colhead{\hd}& \colhead{$\Delta$\hd} & \colhead{A(\hg/\hb)} & \colhead{A(\hg/\hb)}}
\startdata
1 & 1   - 6  & 29 & 1.0  & 1.50 & 0.12 & 0.63 & 0.10& 0.41 & 0.08 & 0.83 & -0.29\\
2 & 6 - 8  & 24 &  1.1   & 1.62 & 0.14 & 0.74 & 0.10 & 0.42 & 0.07 & 0.16 & -0.01\\
3 & 8 - 12  & 31 & 1.34 & 2.66 & 0.13 & 1.07 & 0.10 & 0.49 & 0.08 & 1.08 & 2.31\\
4 & 12-20  & 49 &  2.0  & 2.60 & 0.14 & 1.05 & 0.10 & 0.57 & 0.08 & 1.01 & 1.14\\
5 & 20 - 40 & 25  &  2.5& 3.68 & 0.15 & 1.61 & 0.11 & 0.86 & 0.08 & 0.47 & 0.78\\
6 & 40 - 120 & 16  &3.7 & 5.04 & 0.15 & 2.22 & 0.10 & 1.00 & 0.08 & 0.41 & 1.88\\
\enddata
\tablenotetext{a} {\ The data presented here are for the stacked spectra which exclude individually-identified AGN and composite sources.}
\tablenotetext{b} {\ Line fluxes are in units of $10^{-17}$ ergs s$^{-1}$ cm$^{-2}$. They are not corrected for attenuation or aperture losses.}
\tablenotetext{c} {\ Number of individual spectra used to create the stacked spectrum.}
\tablenotetext{d} {\ Mean continuum signal-to-noise in the stacked spectrum.}
\end{deluxetable}

 \end{document}